\input harvmac


\input epsf

\newcount\figno
\figno=0
\def\fig#1#2#3{
\par\begingroup\parindent=0pt\leftskip=1cm\rightskip=1cm\parindent=0pt
\baselineskip=11pt \global\advance\figno by 1 \midinsert
\epsfxsize=#3 \centerline{\epsfbox{#2}} \vskip 12pt {\bf Figure
\the\figno:} #1\par
\endinsert\endgroup\par
}
\def\figlabel#1{\xdef#1{\the\figno}}

\def\subsubsection#1{\bigskip\noindent
{\it #1}}
\def\subsubsec#1{\subsubsection{#1}}
\def\yboxit#1#2{\vbox{\hrule height #1 \hbox{\vrule width #1
\vbox{#2}\vrule width #1 }\hrule height #1 }}
\def\fillbox#1{\hbox to #1{\vbox to #1{\vfil}\hfil}}
\def\ybox{{\lower 1.3pt \yboxit{0.4pt}{\fillbox{8pt}}\hskip-0.2pt}}
%


\noblackbox

\def\IZ{\relax\ifmmode\mathchoice
{\hbox{\cmss Z\kern-.4em Z}}{\hbox{\cmss Z\kern-.4em Z}}
{\lower.9pt\hbox{\cmsss Z\kern-.4em Z}} {\lower1.2pt\hbox{\cmsss
Z\kern-.4em Z}}\else{\cmss Z\kern-.4em Z}\fi}

\def\p{\partial}

\font\cmss=cmss10 \font\cmsss=cmss10 at 7pt
\def\IR{\relax{\rm I\kern-.18em R}}
\def\inbar{\,\vrule height1.5ex width.4pt depth0pt}
\def\IC{\relax\hbox{$\inbar\kern-.3em{\rm C}$}}
\def\IR{\relax{\rm I\kern-.18em R}}
\def\IP{\relax{\rm I\kern-.18em P}}

\def\frac#1#2{{#1 \over #2}}


\def\journal#1&#2(#3){\unskip, \sl #1\ \bf #2 \rm(19#3) }
\def\andjournal#1&#2(#3){\sl #1~\bf #2 \rm (19#3) }

\def\d{\partial}

%

%
\catcode`\@=11
\def\slash#1{\mathord{\mathpalette\c@ncel{#1}}}
\overfullrule=0pt

\def\HH{{\cal H}}

\def\LL{{\cal L}}

\def\NN{{\cal N}}
\def\OO{{\cal O}}

\def\WW{{\cal W}}
\def\XX{{\cal X}}

\def\ZZ{{\cal Z}}

\def\eps{\epsilon}

\def\underrel#1\over#2{\mathrel{\mathop{\kern\z@#1}\limits_{#2}}}

\catcode`\@=12

\def\({\left(}
\def\){\right)}
\def\[{\left[}
\def\]{\right]}
\def\<{\langle}
\def\>{\rangle}
\def\half{{1\over 2}}
\def\d{\partial}
\def\tr{{\rm Tr}}
\def\|{\biggl|}

\def\exp{{\rm exp}}

\def\cf{{\it c.f.}}

\def\bz{{\bar z}}


%


\lref\DixonQV{
  L.~J.~Dixon, D.~Friedan, E.~J.~Martinec and S.~H.~Shenker,
  ``The Conformal Field Theory Of Orbifolds,''
  Nucl.\ Phys.\ B {\bf 282}, 13 (1987).
}

\lref\LauerAX{
  J.~Lauer, J.~Mas and H.~P.~Nilles,
  ``Duality And The Role Of Nonperturbative Effects On The World Sheet,''
  Phys.\ Lett.\ B {\bf 226}, 251 (1989).
}

\lref\DixonJW{
  L.~J.~Dixon, J.~A.~Harvey, C.~Vafa and E.~Witten,
  ``Strings On Orbifolds,''
  Nucl.\ Phys.\ B {\bf 261}, 678 (1985).
}

\lref\StromingerKU{
  A.~Strominger,
  ``Topology Of Superstring Compactification,''
NSF-ITP-85-109
{\it Presented at Santa Barbara Workshop on Unified String Theories,
Santa Barbara, CA, Jul 29 - Aug 16, 1985} }

\lref\LS{A.~Lawrence and A.~Sever, "Quantum corrections to
Fayet-Iliopoulos D-terms in field theory and string theory", to
appear.}

\lref\GimonRQ{
  E.~G.~Gimon and J.~Polchinski,
  ``Consistency Conditions for Orientifolds and D-Manifolds,''
  Phys.\ Rev.\ D {\bf 54}, 1667 (1996)
  [arXiv:hep-th/9601038].
}

\lref\VafaIH{
  C.~Vafa,
  ``Quantum symmetries of string vacua'',
  Mod.\ Phys.\ Lett.\ A {\bf 4}, 1615 (1989).
}

\lref\DineDK{
  M.~Dine and M.~Graesser,
  ``CPT and other symmetries in string / M theory,''
  JHEP {\bf 0501}, 038 (2005)
  [arXiv:hep-th/0409209].
}

\lref\LustCX{
  D.~Lust, P.~Mayr, R.~Richter and S.~Stieberger,
  ``Scattering of gauge, matter, and moduli fields from intersecting  branes,''
  Nucl.\ Phys.\  B {\bf 696}, 205 (2004)
  [arXiv:hep-th/0404134].
}
\lref\BergJA{
  M.~Berg, M.~Haack and B.~Kors,
  ``String loop corrections to Kaehler potentials in orientifolds,''
  JHEP {\bf 0511}, 030 (2005)
  [arXiv:hep-th/0508043].
}

\lref\AngelantonjUY{
  C.~Angelantonj, M.~Bianchi, G.~Pradisi, A.~Sagnotti and Y.~S.~Stanev,
  ``Chiral asymmetry in four-dimensional open- string vacua,''
  Phys.\ Lett.\  B {\bf 385}, 96 (1996)
  [arXiv:hep-th/9606169].
}

\lref\FriedanGE{
  D.~Friedan, E.~J.~Martinec and S.~H.~Shenker,
  ``Conformal Invariance, Supersymmetry And String Theory,''
  Nucl.\ Phys.\  B {\bf 271}, 93 (1986).
}
\lref\PolchinskiRQ{
  J.~Polchinski,
  ``String theory. Vol. 1: An introduction to the bosonic string,''
{\it  Cambridge, UK: Univ. Pr. (1998) 402 p};
 and
  ``String theory. Vol. 2: Superstring theory and beyond,''
{\it  Cambridge, UK: Univ. Pr. (1998) 531 p} }
\lref\DouglasSW{
  M.~R.~Douglas and G.~W.~Moore,
  ``D-branes, Quivers, and ALE Instantons,''
  arXiv:hep-th/9603167.
}
\lref\OoguriCK{
  H.~Ooguri, Y.~Oz and Z.~Yin,
  ``D-branes on Calabi-Yau spaces and their mirrors,''
  Nucl.\ Phys.\  B {\bf 477}, 407 (1996)
  [arXiv:hep-th/9606112].
}
\lref\PoppitzDJ{
  E.~Poppitz,
  ``On the one loop Fayet-Iliopoulos term in chiral four dimensional type I
  orbifolds,''
  Nucl.\ Phys.\  B {\bf 542}, 31 (1999)
  [arXiv:hep-th/9810010].
}
\lref\AtickGY{
  J.~J.~Atick, L.~J.~Dixon and A.~Sen,
  ``String Calculation Of Fayet-Iliopoulos D Terms In Arbitrary Supersymmetric
  Compactifications,''
  Nucl.\ Phys.\  B {\bf 292}, 109 (1987).
}
\lref\DineGJ{
  M.~Dine, I.~Ichinose and N.~Seiberg,
  ``F Terms And D Terms In String Theory,''
  Nucl.\ Phys.\  B {\bf 293}, 253 (1987).
}
\lref\LawrenceSM{
  A.~Lawrence and J.~McGreevy,
  ``D-terms and D-strings in open string models,''
  JHEP {\bf 0410}, 056 (2004)
  [arXiv:hep-th/0409284].
}
\lref\KachruYS{
  S.~Kachru and E.~Silverstein,
  ``4d conformal theories and strings on orbifolds,''
  Phys.\ Rev.\ Lett.\  {\bf 80}, 4855 (1998)
  [arXiv:hep-th/9802183].
}
\lref\LawrenceJA{
  A.~E.~Lawrence, N.~Nekrasov and C.~Vafa,
  ``On conformal field theories in four dimensions,''
  Nucl.\ Phys.\  B {\bf 533}, 199 (1998)
  [arXiv:hep-th/9803015].
}

\lref\BainFB{
  P.~Bain and M.~Berg,
  ``Effective action of matter fields in four-dimensional string
  orientifolds,''
  JHEP {\bf 0004}, 013 (2000)
  [arXiv:hep-th/0003185].
}
\lref\BachasZT{
  C.~Bachas and C.~Fabre,
  ``Threshold Effects in Open-String Theory,''
  Nucl.\ Phys.\  B {\bf 476}, 418 (1996)
  [arXiv:hep-th/9605028].
}
\lref\AntoniadisGE{
  I.~Antoniadis, C.~Bachas and E.~Dudas,
  ``Gauge couplings in four-dimensional type I string orbifolds,''
  Nucl.\ Phys.\  B {\bf 560}, 93 (1999)
  [arXiv:hep-th/9906039].
}
\lref\AkerblomUC{
  N.~Akerblom, R.~Blumenhagen, D.~Lust and M.~Schmidt-Sommerfeld,
  ``Instantons and Holomorphic Couplings in Intersecting D-brane Models,''
  arXiv:0705.2366 [hep-th].
}
\lref\FischlerZK{
  W.~Fischler, H.~P.~Nilles, J.~Polchinski, S.~Raby and L.~Susskind,
  ``Vanishing Renormalization Of The D Term In Supersymmetric U(1) Theories,''
  Phys.\ Rev.\ Lett.\  {\bf 47}, 757 (1981).
}

\lref\KlebanovMY{
  I.~R.~Klebanov and E.~Witten,
  ``Proton decay in intersecting D-brane models,''
  Nucl.\ Phys.\  B {\bf 664}, 3 (2003)
  [arXiv:hep-th/0304079].
}
\lref\DvaliZH{
  G.~Dvali, R.~Kallosh and A.~Van Proeyen,
  ``D-term strings,''
  JHEP {\bf 0401}, 035 (2004)
  [arXiv:hep-th/0312005].
}
\lref\BinetruyHH{
  P.~Binetruy, G.~Dvali, R.~Kallosh and A.~Van Proeyen,
  ``Fayet-Iliopoulos terms in supergravity and cosmology,''
  Class.\ Quant.\ Grav.\  {\bf 21}, 3137 (2004)
  [arXiv:hep-th/0402046].
}
\lref\SvrcekYI{
  P.~Svrcek and E.~Witten,
  ``Axions in string theory,''
  JHEP {\bf 0606}, 051 (2006)
  [arXiv:hep-th/0605206].
}

\lref\BertoliniQH{
  M.~Bertolini, M.~Billo, A.~Lerda, J.~F.~Morales and R.~Russo,
  ``Brane world effective actions for D-branes with fluxes,''
  Nucl.\ Phys.\  B {\bf 743}, 1 (2006)
  [arXiv:hep-th/0512067].
}

\Title{\vbox{\baselineskip12pt \hbox{BRX TH-584,~KITP-06-126}}}
{\vbox{ \centerline{Scattering of twist fields}
\smallskip
\smallskip\centerline{from D-branes and orientifolds}
\smallskip
 }} \centerline{Albion Lawrence and Amit Sever}
\bigskip
\centerline{{\it Brandeis Theory Group, Martin Fisher School of
Physics,}} \centerline{{\it Brandeis University, Waltham, MA
02454-9110}}

\bigskip
\bigskip
\noindent

We compute the two-point function for $Z_N$ orbifold twist fields on
the disk and $RP^2$. We apply this to a computation of the
$\CO(g_s)$ correction to the K\"ahler potential for (the symmetric
combination of) blow-up modes in type I string theory on
$T^6/\IZ_3$. This is related by supersymmetry to the one-loop field
dependent correction to the Fayet-Iliopoulos D-term for the
anomalous $U(1)$ factor. We find this correction to be non-vanishing
away from the orbifold point.

\Date{June 2007}


\newsec{Introduction}

In this work we will consider the scattering of orbifold blow-up
modes off of D-branes and orientifold planes.  Our motivation is the
study of the effective action of 4d models with open strings and
low-energy $\NN=1$ spacetime supersymmetry. Orbifold blow-up modes
are typically moduli or light scalars in the problem, and control
some of the coupling constants in the open string sector.

More precisely, we will compute the two-point function of orbifold
twist fields on the disk and $RP^2$. Among other things, this allows
us to compute the $\CO(g_s)$ correction to the quadratic part of the
K\"ahler potential for the blow-up modes. To the best of our
knowledge such computations have not been done for twisted sector
moduli -- most of the focus has been on corrections to the K\"ahler
potential for the untwisted moduli, as in
\refs{\LustCX,\BergJA}.\foot{See \BertoliniQH\ for a generalization
of \LustCX.} For twisted moduli, this term is important for
computing the Fayet-Iliopoulos D-terms in these models, as we will
discuss in \S4.

The difficulty in this calculation stems from the correlators for
the bosonic twist fields. For the CFT of a free boson or free
fermion, one may use the "doubling trick". This trick constructs the
disk or real projective plane from an identification on the complex
plane, and uses this identification to relate the antiholomorphic
part of the operators on the disk or real projective plane to a
holomorphic part on the complex plane.  Thus all correlators can be
transformed to holomorphic correlators on the sphere. However, this
trick requires that one can factorize the operators into holomorphic
and antiholomorphic pieces. For bosonic twist fields, this is not
possible, as can be seen from the calculation of the four-point
function of twist operators on the sphere \refs{\DixonQV}.

We will compute the two-point function of twist fields on the disk
and on $RP^2$ for a variety of boundary conditions, using the
"stress tensor method" in \refs{\DixonQV}. Our calculation will
amount to performing the doubling trick on the additional insertions
in \refs{\DixonQV}\ and mapping all of the monodromy integrals to
holomorphic integrals on the sphere. The nonfactorizability of the
bosonic twist fields will arise from the non-locality of the
monodromy conditions on the sphere. The result is very close to the
calculation of the 4-point function on the sphere.

We then apply this to a calculation of the $\OO(g_s)$ correction to
the spacetime kinetic term for twisted sector moduli in the type I
$T^6/Z_3$ orbifold \AngelantonjUY. After canceling all of the
tadpoles and unphysical divergences, the different contributions to
the amplitude (coming from the disk and the projective plane) join
at the boundary of their moduli spaces into a single integral over
the modular parameter of four points on the sphere, leading to a
finite non-zero answer. We are unable to evaluate that integral
analytically, or even numerically for finite volume of the torus;
however, we have done the integral numerically in the limit of large
volume of the orbifold, and found it to be non-zero.\foot{Ref.
\refs{\AkerblomUC}, which appeared as this paper was being
completed, has also argued that the one-loop correction to the FI
D-term is nonvanishing away from the orientifold point in type IIA
models on $T^6$ with intersecting 6-branes.}

The outline of this paper is as follows. In \S2\ we will compute the
two-point-function of bosonic twist fields on the disk and real
projective plane.  In \S3\ we will compute the correlation function
of fermionic twist fields.  In \S4\ we will combine these to
calculate the two-point function of the diagonal sum of the
twisted sector moduli of type I string theory compactified on
$T^6/Z_3$.  We study the
divergence structure of these amplitudes, discuss their spacetime
interpretation,  and extract the one-loop correction to the K\"ahler potential.
In \S5\ we will conclude with some comments on the
spacetime physics of this calculation. The Appendices collect some
facts about hypergeometric functions.

\newsec{Bosonic correlation functions}

In this section we wish to compute the two-point function of bosonic
$\IZ_N$ twist fields on worldsheets with disk and $RP^2$ topology,
for the case of $T^2/\IZ_N$ and $\IC/\IZ_N$.  The $T^2$ and $\IC$
are described by a complex
coordinate $X$ and $Z_N$ is generated by the action
$X \to X e^{2\pi i/N}$.

We will use the stress tensor method for this calculation, as
developed in \DixonQV. The essence of this method is as follows.
In order to compute the two-point function of two twist
fields  $\sigma_1(z_1,\bar z_1)$,$\sigma_2(z_2,\bar z_2)$, we
compute this two-point function with the insertion $\d X(z) \d
\bar{X}(w)$. The monodromy conditions on $\d X,\d\bar{X}$ determine
this correlator. The leading nonsingular part of this composite
operator as $z\to w$ is the stress tensor, and we can then integrate
the conformal Ward identities to extract the original correlator.

Although we cannot apply the doubling trick directly to the twist
fields, we {\it can} apply it to the $\d X(z) \d \bar{X}(w)$
insertion.  This application, and the modification of the monodromy
conditions to the disk and $RP^2$ topology, are the essential
innovations of this section.

\subsec{Boundary conditions}

Boundary conditions on $X$ can be imposed by considering $X$ as a
function on the complex plane, and then projecting onto
configurations invariant under the $\IZ_2$ action
\eqn\discretesym{
    z \to \epsilon_2/\bar{z}~.}
The fundamental domain of this map is the unit disk $|z|\le 1$. It
has the topology of a disk for $\epsilon_2=1$ and of $RP^2$ for
$\epsilon_2=-1$. The action of this symmetry on $X$ is:
\eqn\modding{X(z,\bar z)=\epsilon_1X\(\epsilon_2/\bar
z,\epsilon_2/z\)~.}
Here $\epsilon_1=1\ (-1)$ for the coordinates parallel
(perpendicular) to the brane/orientifold plane, and we choose the
brane/orientifold to be located at $X = 0$ in the directions
perpendicular to it.

Using \modding, the holomorphic and anti-holomorphic fields -- $\d X$
and $\bar\d X$, respectively~--  in the fundamental region can be combined into a
single holomorphic function on the sphere by defining:
\eqn\zbarzrel{\d X(z)=\epsilon_1\bar\d X(\bar u)\({\d \bar u\over\d
z}\)^{(\Delta = 1)}\|_{\bar
u=\epsilon_2/z}=-\epsilon_1\epsilon_2\bar\d X(\epsilon_2/z)/z^2}
for $|z|>1$, or equivalently
\eqn\zbarzreltwo{
z\d X(z)=-\epsilon_1\bar u\bar\d X(\bar u)\|_{\bar u=\epsilon_2/z}~.}
Note that the $\IZ_2$ \discretesym\ acts on $\d X$ with the Jacobian
factor expected of an operator with dimension $\Delta = 1$.

We are interested in the correlators
\eqn\stinsert{
    C = \langle \p X(z) \p \bar{X}(w)\sigma_1(z_1,\bar{z}_1)
    \sigma_2(z_2,\bar{z}_2)\rangle~,}
where $\sigma_{1,2}$ are bosonic twist fields.  As functions of
$z,w$ these satisfy additional boundary conditions as $z,w \to
z_{1,2}$.  The correlators will be further constrained by the
interaction of these boundary conditions with \zbarzrel.

More precisely, consider a bosonic twist operator $\sigma_+$ that
twists the field $X(z,\bar z)$ by $e^{2\pi ik/N}$ at $(w,\bar w)$.
The boson $X$ has the following OPEs with $\sigma_+$
\refs{\DixonQV}:
\eqn\bosonopes{ \eqalign{\d X(z)\sigma_+(w,\bar w)\sim
&(z-w)^{-(1-k/N)}\tau_+(w,\bar w)\cr \bar\d X(\bar z)\sigma_+(w,\bar
w)\sim &(\bar z-\bar w)^{-k/N}\tilde\tau'_+(w,\bar w)\cr
\d\bar{X}(z) \sigma_+(w,\bar{w})\sim & (z - w)^{-k/N}
\tau'_+(w,\bar{w})\cr \bar{\d}\bar{X}(z) \sigma_+(w,\bar w) \sim&
(\bar z - \bar w)^{-(1-k/N)} \tilde\tau_+(w,\bar w)~.}}
Here $\tau,\tilde\tau,\tau',\tilde\tau'$ are known as excited twist
fields. We also define the "conjugate twist field" $\sigma_-$ as
having the OPEs \bosonopes\ with $k/N \leftrightarrow 1-k/N$, and
$\tau_+ \leftrightarrow \tau_-$, etc.

\subsec{Correlation functions and stress tensor insertions}

In \S4, we will find that we need to compute the following two-point
functions:
\eqn\twopoint{
    H^{\eps_1,\eps_2}_{-\pm}
    = \langle \sigma_{-}(z_1,\bz_1) \sigma_{\pm}(z_2,\bz_2)
    \rangle~.}
The quantum symmetry that would set $H_{--} = 0$ on the sphere is
broken on the disk and on $RP^2$.  We will give a spacetime argument
in \S4, but for now we note that we find no reason for
it to vanish from a two-dimensional point of view, either.

The first step in computing $H$ will be to compute the following
connected correlators:
\eqn\stinsert{ \eqalign{
    g^{\eps_1,\eps_2}_{-\pm}(z) & = \frac{\langle -\half \p X(z) \p\bar{X}(z) \sigma_{-}(z_1,\bz_1)
        \sigma_{\pm}(z_2,\bz_2) \rangle}
        {\langle \sigma_{-}(z_1,\bz_1) \sigma_{\pm} (z_2,\bz_2)}\cr
    h^{\eps_1,\eps_2}_{-\pm}(z,\bz) & =
        \frac{\langle -\half \bar{\p} X(\bz) \p\bar{X}(z) \sigma_{-}(z_1,\bz_1)
        \sigma_{\pm}(z_2,\bz_2) \rangle}
        {\langle \sigma_{-}(z_1,\bz_1) \sigma_{\pm}
        (z_2,\bz_2)\rangle}~.}}
These are determined by imposing the boundary conditions and
monodromy conditions on the $z$-dependence of these correlators, as
in \DixonQV.

The monodromy conditions arise from the fact that as one takes $X$
around noncontractible contours on the worldsheet, which surround
the same amount of $Z_N$ twist as anti-twist,
$X$ should return to itself up to an element of the toroidal orbifold group (which define
the torus.) Figures 1-3 show
some of the contours which contribute to the
monodromy conditions on the disk for $H_{-+}$, for $H_{--}$ with
$\epsilon_1 = 1$, and for $H_{--}$ with $\epsilon_1 = -1$,
respectively. Figure 4 shows such contours for $RP^2$, for $H_{-+}$.
The contours inside the disk are noncontractible contours, with the
ends at $|z|=1$ identified via the action \discretesym. The full
closed loops are the images of these contours under \discretesym.

We will find that it is enough to impose a single monodromy
condition for each case. The contours that contribute nontrivially
to the monodromy conditions for $\eps_1 = \pm 1$, will be trivially
satisfied or will give equivalent monodromy conditions for $\eps_1 =
\mp 1$. This is unlike the calculation in \refs{\DixonQV} of the
four-point function on the sphere, for which there are two
independent monodromy conditions that must be enforced.

The monodromy conditions on $X$ inside correlation functions
$H_{-\pm}$ are:
\eqn\monodcondit{
    \int_C \left( dz \p X + d\bz\bar{\p} X \right) =
    v~,}
where $v$ is an element of some coset of the Narain lattice
$\Lambda$ \refs{\DixonQV}; we will discuss this further in \S2.4. In
our case there is a further simplification -- the boundary
conditions \modding\ allow one to take the contour integrals above,
and write them as {\it holomorphic} contours in the full complex
plane. These contours are also shown in Figs. 1-4.

In performing the path integral to compute $H$, we sum over the
different possible monodromies, as we will discuss in \S2.4. This
sum can be simplified as follows. For a given monodromy
\monodcondit, we split $X$ into two pieces $X = X_{cl} + X_q$. The
first is the "classical" piece, it satisfies the classical equations
of motion $\d\bar\d X_{cl} = 0$
and the conditions \monodcondit\ by
itself. The second piece $X_q$ is the "quantum" piece containing the
fluctuations about the classical solution, and solves \monodcondit\
with $v = 0$. The full correlator can be written as
\eqn\fullcorrelator{\eqalign{
    H^{\eps_1,\eps_2}_{-\pm} & = H^{\eps_1,\eps_2}_{q;-\pm} H^{\eps_1,\eps_2}_{cl;-\pm} \cr
    H^{\eps_1,\eps_2}_{cl;-\pm}
    & = \sum_v e^{-S^{\eps_1,\eps_2}_{cl,-\pm}[v]}~,}}
where in $H_{cl}$ the sum is over the elements in the corresponding
coset lattice \monodcondit.

The connected correlation functions $g_{\mp,+}$ are independent of
the classical piece.  They are determined by the boundary
conditions, together with the monodromy conditions \monodcondit\ for
$v = 0$. Once $g$ is determined, one can use the conformal Ward
identities to compute $H_q$. We will give an outline of and results
of this computation in \S2.3.  The sum over classical solutions will
be described in \S2.4.

\subsec{The quantum piece}

We will compute $H^{\eps_1,\eps_2}_{q; -\pm}$ case by case. Before
presenting the answers we will sketch the basic strategy.

To construct $H_q$ we need first to compute $g_{-\pm}$. Because $g$
is a connected correlator, any dependence on the explicit factor of $\p X_{cl}$ vanishes,
and it can be taken as the connected correlator of $\d X_q(z) \d\bar{X}_q(w)$.

The asymptotic conditions as $z$ approaches any of the other insertions in \stinsert,
together with the condition \zbarzrel, determines $g_{\mp,+}(z,w)$ up to some function
$A_\pm(z_1,\bar z_1,z_2,\bar z_2)$:
\eqn\bcsolution{
    \eqalign{g_{-+}(z,w)=\omega_k(z)\omega_{N-k}(w)\{&\frac{k}{N}{(z-z_1)(z-1/{\bar
z_2})(w-z_2)(w-1/{\bar z_1})\over
(z-w)^2}\cr+&\(1-\frac{k}N\){(z-z_2)(z-1/{\bar
z_1})(w-z_1)(w-1/{\bar z_2})\over (z-w)^2}+A_{+}\}~,} }
where
\eqn\classicalsoln{
    \omega_k(z)\equiv[(z-z_1)(z-\epsilon_2/\bar z_2)]^{-k/N}
    [(z-z_2)(z-\epsilon_2/\bar z_1)]^{-(1-k/N)}~.
}
Similarly, $g_{--}$ is obtained from $g_{-+}$ by exchanging $z_2
\leftrightarrow \epsilon_2/\bar z_2$, and $A_+\leftrightarrow A_-$.

$A$ is determined by the monodromy condition for $X_{qu}$, which
is \monodcondit\ with $v=0$ for an appropriate choice of contour $C$.
We will call these contours closed if their image on the branched
cover of the sphere, as given by \discretesym,
is closed.

Consider the space of all closed contours in $|z| \leq 1$, up to
homotopy. In our case this space is one-dimensional, with a single
contour $\LL_0$ as a basis element. (In the four-point function on
the sphere, the space is two-dimensional \refs{\DixonQV}). By using
the map \discretesym, the monodromy integral \monodcondit\ over the
contour $\LL_0$ in the fundamental region can be mapped to an
integral over a single closed loop $\gamma$ on the sphere (the
covering space), times a numerical factor $1/k$, where
$k=(3-\epsilon_1\epsilon_2)/2$ (this will be unimportant for the
monodromy condition for $X_q$, but important for $X_{cl}$). The map
\discretesym\ maps a twist operator to an anti-twist operator at the
image point: as a result, the total amount of twist encircled by the
loop $\gamma$ on the sphere is zero (and one may think of the
monodromy condition for $X_{qu}$ as a monodromy condition on the
sphere.) The fundamental closed loop on the sphere $\gamma$ is drawn
for each case in figures 1-4.  In these figures, the original
contour $C$ inside the fundamental region $|z|\leq 1$ is black,
while the remainder $\gamma - C$ outside the fundamental region is
blue.
%
The monodromy condition for $X_{qu}$ can be written as:
\eqn\monodromy{\Delta_{\LL_0} X_{qu}={1\over k}\oint_\gamma dz\d
X_{qu}+{1\over k}\oint_\gamma d\bar z\bar\d X_{qu}=0~.}
Next, we can turn the left hand side of  \monodromy\ into a single
holomorphic integral.  Using \discretesym, the condition \monodcondit\ for $X_q$
becomes:
\eqn\holomonod{
    \oint_\gamma dz \d X_{qu}+\epsilon_1\oint_{\bar\gamma} dz \d
X_{qu}=0~,}
where $\bar\gamma$ (also shown in figures 1-4) is the image of $\gamma$ under
\discretesym. Finally, we can apply \holomonod\ to $g_{-\pm}$ to find that:
\eqn\looprel{\oint_\gamma dz g_{-\pm}+\epsilon_1\oint_{\bar\gamma}
dz g_{-\pm}=0~.}
This equation can be solved for $A_{\pm}$, and so determines $g_{\pm}$.

Given $g_{-\pm}$, we then use the conformal Ward identities to
extract and solve a differential equation for $H_{q;-\pm}$. First,
we can use the $SL(2,\IR)$ symmetry of correlators on the disk and
$RP^2$ to set $(z_1,z_2)=(0,y)$, where $y\in[0,1]$. Taking
$z\to w$ we have
\eqn\confwi{\eqalign{&{\<T(z)\sigma_-(0,0)\sigma_\pm(y,\bar{y})\>\over
\<\sigma_-(0,0)\sigma_\pm(y,\bar{y}\>}\biggl|_{\bar{y}=y}=\lim_{z\to
w}\[g_\pm(z,w)-\frac1{(z-w)^2}\]\cr\cr
=&\half\frac{k}N\(1-\frac{k}N\)\(\frac1z\pm\frac1{z-\epsilon_2/y}\mp\frac1{z-y}\)^2
+{\widetilde A\over z(z-\epsilon_2/y)(z-y)}~,}}
where
\eqn\rescaled{\widetilde A_{\pm}(y)\equiv\lim_{z_1\to
0}-\epsilon_2\bar z_1A_{\pm}(z_1,y,\epsilon_2/y,\epsilon_2/\bar
z_1)~.}
As $z\to y$ we have the operator product expansion
\eqn\stope{T(z)\sigma_\pm(y)\sim
{h_\sigma\sigma_\pm(y)\over(z-y)^2}+{\d\sigma_\pm(y)\over(z-y)}+\dots~.}
Integrating along a small contour around $y$ and noting that $H_{q}$
is a function of $y\bar y$ only (we have used the $SL(2,\IR)$
symmetry to set $y=\bar y$), leads to
\eqn\Zeq{\eqalign{\[\d_y\ln H_{q;-\pm}^{\eps_1\eps_2}(y\bar
y)\]\|_{y=\bar y}=&\half\d_y\ln H_{q;-\pm}^{\eps_1,\eps_2}(y^2)\cr
=&\ \mp\frac{k}N\(1-\frac{k}N\)\(\frac1y\pm\frac1{y-\epsilon_2/y}\)+
{\widetilde A_{\pm}\over y(y-\epsilon_2/y)}~.}}
%
After solving for $H_{q;-pm}^{\eps_1,\eps_2}(y^2)$, we will rewrite
it as a function of the cross-ratio on the sphere
\eqn\xxx{x\equiv\[{(z_1-z_2)(z_3-z_4)\over
(z_1-z_3)(z_2-z_4)}\]^{\pm 1}
\|_{(z_1,z_2,z_3,z_4)=(0,y,\epsilon_2/y,\infty)}=\epsilon_2 y^{\pm
2}~,}
where the sign of the exponent corresponds to the sign of the twist
field inserted at $(z_2,\bar{z}_2)$. Let us now consider each case in
turn.

\subsubsec{The disk ($\eps_2 = 1$)}

For each value of $\epsilon_1$ on the disk, there is a single nontrivial closed
contour allowed by the boundary conditions. For $\epsilon_{1} = 1$,
these contours and their images under \discretesym\ are shown in
Figures 1a (for $g_{-+}$) and 2 (for $g_{--}$).  For $\epsilon_1 =
-1$ the contours are shown in Figure 1b (for $g_{-+}$) and 3 (for
$g_{--}$). We can insert these contours into \looprel\ to find the
equation for $A_{\pm}$.
\fig{Noncontractible contours on the disk for $H_{-+}$. The interior
of the dashed circle is the fundamental domain of eq. \discretesym.
The black contours inside the disk are noncontractible due to the
boundary conditions; figures (a) and (b) shows the noncontractible
loops which contribute to the monodromy conditions for $\epsilon_1 =
1$, $-1$ respectively. The blue contours are the images of the black
contours under \discretesym.}{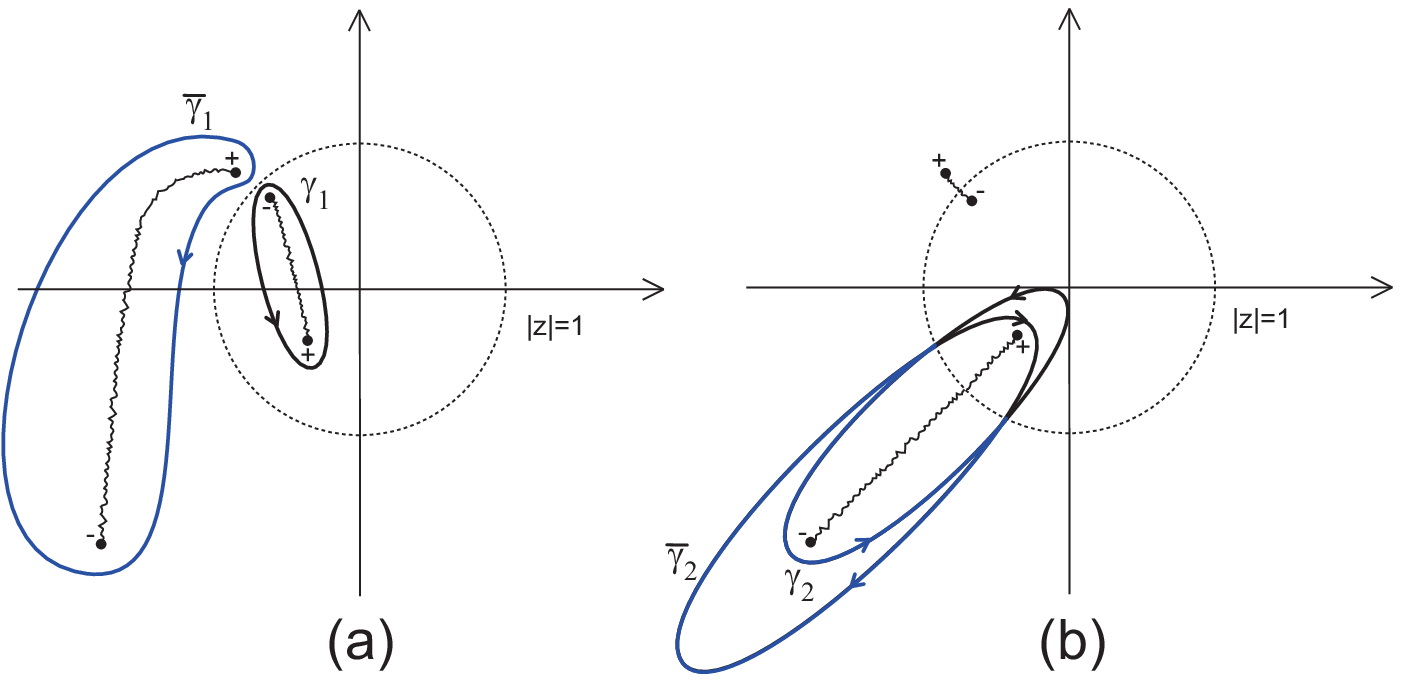}{16
truecm}\figlabel\diskloops
\fig{Noncontractible contours on the disk for $H_{--}$, which
contribute to the monodromy conditions when $\epsilon_1 = 1$.
Figures (a) and (b) are topologically equivalent
contours.}{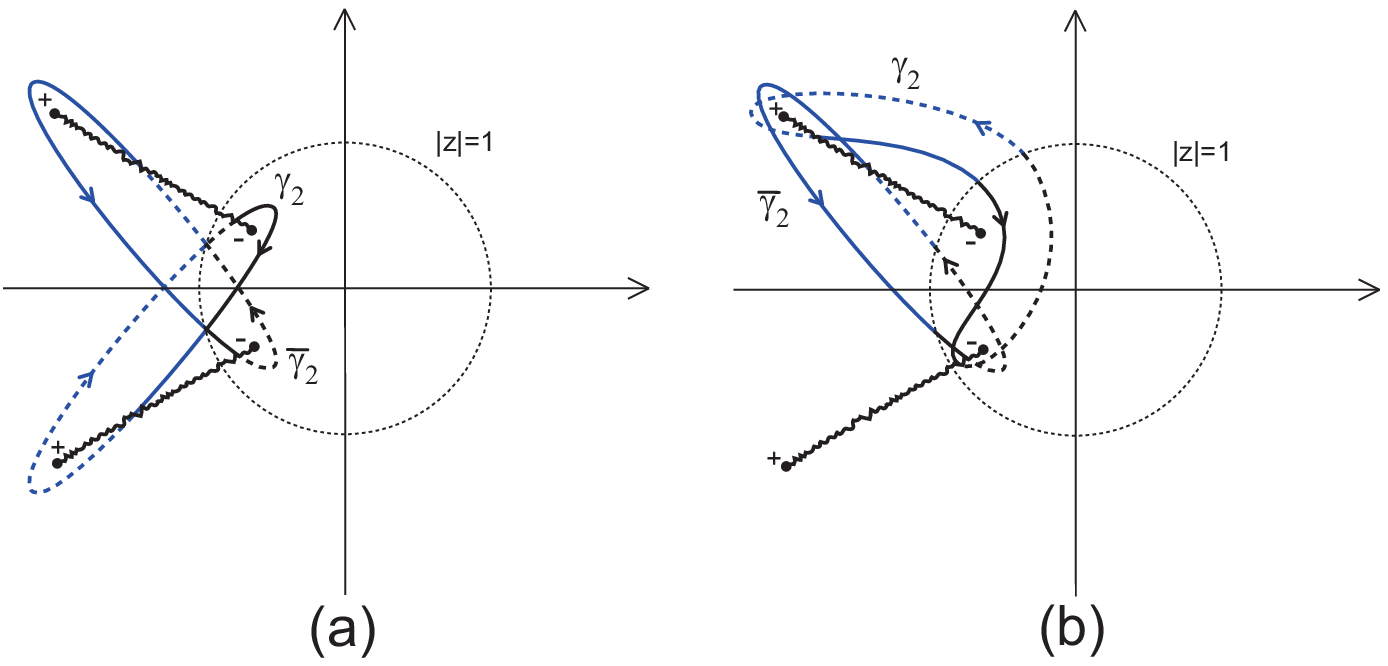} {16 truecm}\figlabel\diskppneumloop
\fig{Noncontractible contours on the disk for $H_{--}$, which
contribute to the monodromy conditions when $\epsilon_1 = -1$.
Figures (a) and (b) are topologically equivalent
contours.}{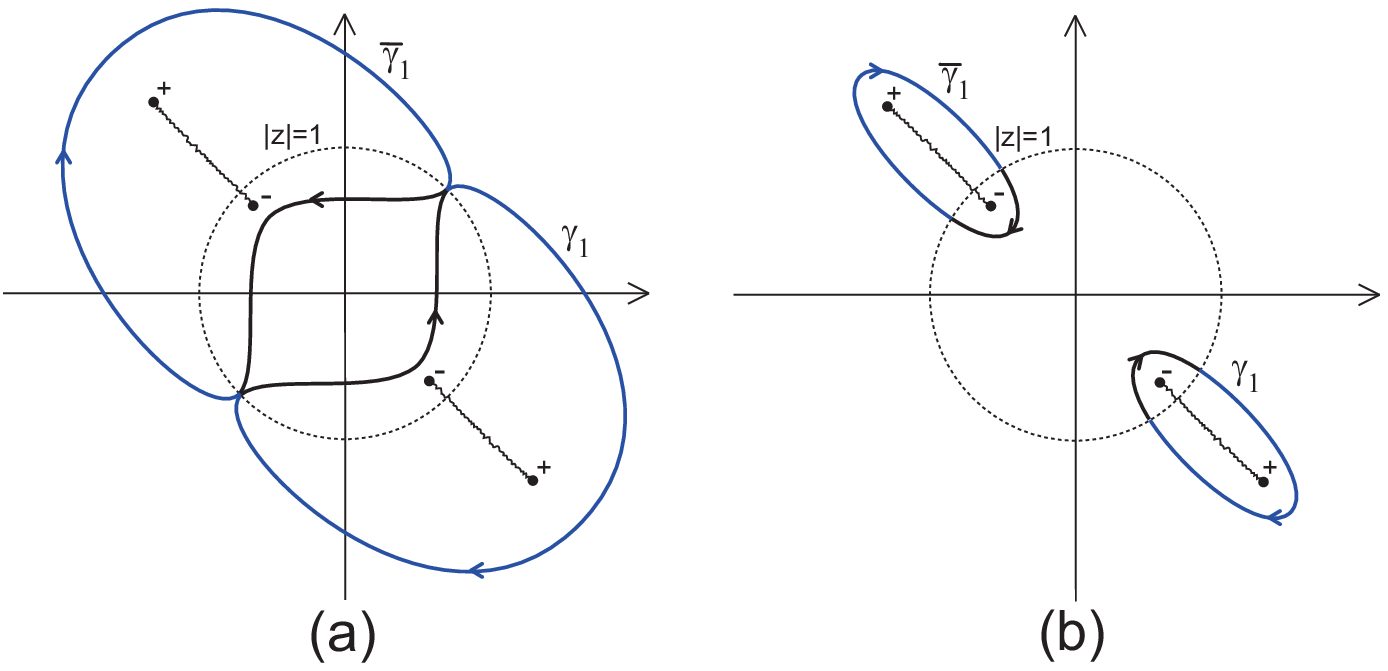} {16 truecm}\figlabel\diskppdirloop

To see that only one of these two loops on the sphere contribute for each correlator,
note that by deforming the loops around infinity in the complex
plane, \foot{Note that there is no simple
pole at infinity.} one can show that in figures 1 and 3, $\bar\gamma_{1}=\eps_1\gamma_{1}$,
while in figure 2, $\bar{\gamma}_2 - \gamma_2$ is proportional to $\gamma_1$
in figure 3 plus the image of the same contour on the second sheet
of the cut plane. The result is that when
$\epsilon_1 = -1$ ($+1$), the monodromy integrals \monodcondit\ for
the contours in Fig 1a and 2 (1b and 3) on the disk do not impose
new conditions.
Therefore, we can simplify the monodromy integrals to
\eqn\gintd{\oint_{\gamma_i} dz g=0~,}
where for $g_{-\pm}$, $i = 1(2)$ for $\eps_1 = \pm 1(\mp 1)$.

We perform the contour integrals with the sole purpose of finding
$A_{\pm}$. Since $A$ is a function of $(z_i,\bz_i)$ only, we can set
$w$ in $g(z,w)$ to be what we like. Therefore, we divide the monodromy
condition by $z_1^{-k/N} \omega_{N-k}$ and send $w\to\infty$. After
performing the resulting integrals (with the help of the identities
in Appendix B) we find that:
\eqn\finaldqu{\eqalign{H_{q;-+}^{+,+}(y)=&{\rm
const}[y^2(1-y^2)]^{-k/N(1-k/N)}\widetilde{F}\(y^2\)^{-1}\cr =&{\rm
const}[x(1-x)]^{-k/N(1-k/N)}\widetilde{F}\(x\)^{-1}\cr
H_{q;--}^{+,+}(y)=&{\rm const}(1-y^{-2})^{-k/N(1-k/N)}\widetilde F
\(y^{-2}\)^{-1}\cr =&{\rm const}(1-x)^{-k/N(1-k/N)}\widetilde
F\(x\)^{-1} \cr H_{q;-+}^{-,+}(y)=&{\rm
const}[y^2(1-y^2)]^{-k/N(1-k/N)}F\(1-y^2\)^{-1}\cr =&{\rm
const}[x(1-x)]^{-k/N(1-k/N)}F\(1-x\)^{-1} \cr
H_{q;--}^{-,+}(y)=&{\rm
const}(1-y^{-2})^{-k/N(1-k/N)}F\(1-y^{-2}\)^{-1}\cr =&{\rm
const}(1-x)^{-k/N(1-k/N)}F\(1-x\)^{-1}~,}}
where $x$ is defined in \xxx, and $\widetilde F(x)= \widetilde
F(1-k/N,k/N,1;x)$ is defined in Appendix $A$.

\subsubsec{The projective plane ($\eps_2 = -1$)}

We proceed as in the case of the disk. As before, for $\epsilon_1=1$
we find that the sole relevant monodromy condition is:
\eqn\gpint{\oint_{\gamma_1} dz g_{-+}=0~,}
where $\gamma_1$ is drown in Figure 4a for $g_{-+}$, and we have
found that it is equal to $\bar{\gamma}_1$ by deforming the loop
around infinity in the complex plane. Similarly, by deforming the
loop $\bar\gamma_2$ in Fig 4b (see Fig 5), we find that:
\eqn\gpintdef{
\oint_{\bar\gamma_2} dz g=-\oint_{\gamma_2} dz g+(1-e^{2\pi ik/N})\oint_{\gamma_1} dz
g~,}
so that for $\epsilon_1=1$, the monodromy condition for the loop
$\gamma_2$ reads:
\eqn\gpintm{\oint_{\gamma_2} dz g+\oint_{\bar\gamma_2} dz
g=(1-e^{2\pi ik/N})\oint_{\gamma_1} dz g=0}
and is therefore equivalent to the monodromy condition obtained from
the loop $\gamma_1$.
%
\fig{Noncontractible loops for $g_{-+}$ on $RP^2$.  Figures (a) and
(b) are for $\epsilon_1 = +1$ and $-1$,
respectively.}{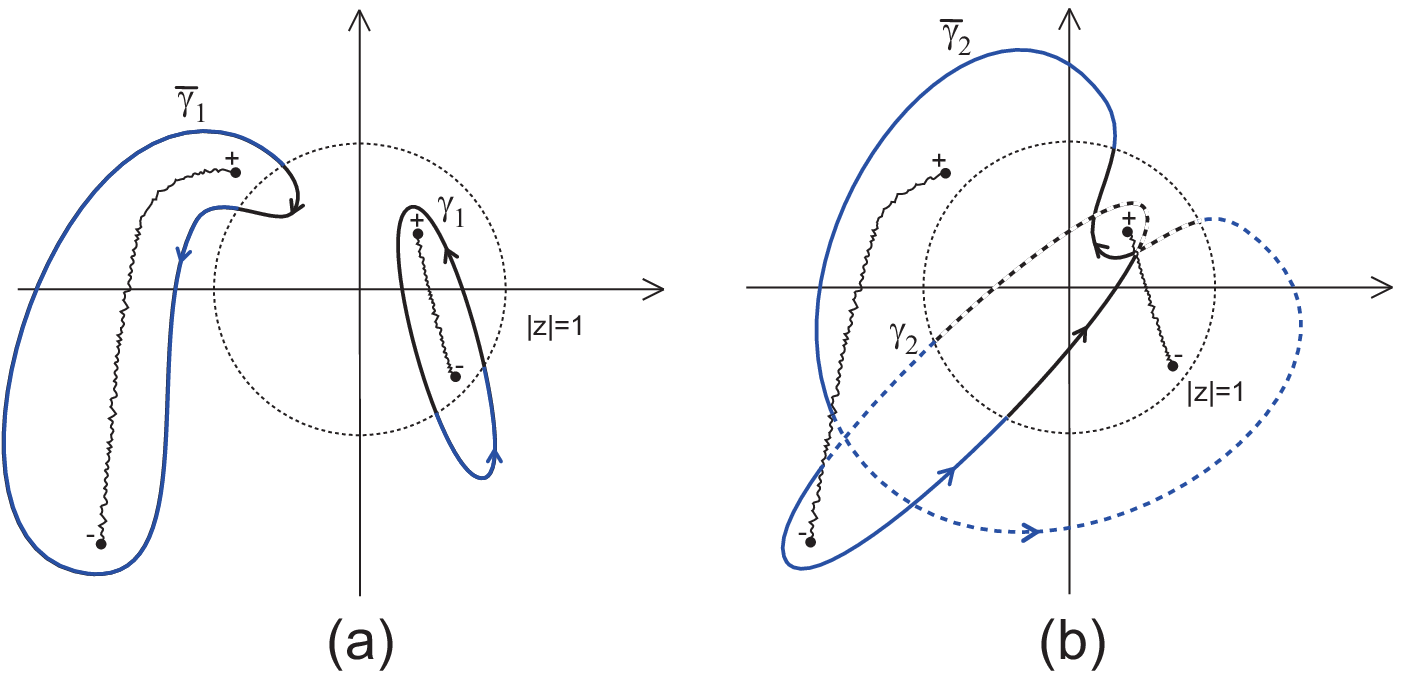}{16 truecm} \figlabel\rponeloops

For $\epsilon_1=-1$, the monodromy condition obtained from the loop
$\gamma_1$ is trivially satisfied (as $\gamma_1=\bar{\gamma}_1$),
whereas the monodromy condition obtained from the loop $\gamma_2$ is
non-trivial.\foot{For $H^{+-}_{--}$ one may equivalently consider
the loop in Fig 3a.}

As before, we divide the monodromy condition (\gpint\ or \gpintm) by
$z_1^{-k/N}\omega_{N-k}(w)$, send $w\to\infty$ and use the
$SL(2,\IR)$ symmetry to set set $(z_1,z_2)=(0,y)$, where
$y\in[0,1]$. The results of the calculation are:
\eqn\finalrponeq{\eqalign{ H_{q;-+}^{+,-}(y)=&{\rm
const}[y^2(1+y^2)]^{-k/N(1-k/N)}F\(-y^2\)^{-1}\cr =&{\rm
const}[x(x-1)]^{-k/N(1-k/N)}F\(x\)^{-1}\cr H_{q;--}^{+,-}(y)=&{\rm
const}(1+y^{-2})^{-k/N(1-k/N)}F\(-y^{-2}\)^{-1}\cr =&{\rm
const}(x-1)^{-k/N(1-k/N)}F\(x\)^{-1}\cr H_{-+\ qu}^{-,-}(y)=&{\rm
const}[y^2(1+y^2)]^{-k/N(1-k/N)}\widetilde F\(1+y^2\)^{-1}\cr =&{\rm
const}[x(x-1)]^{-k/N(1-k/N)}\widetilde F\(1-x\)^{-1}\cr
H_{q;--}^{-,-}(y)=&{\rm const}(1+y^{-2})^{-k/N(1-k/N)}\widetilde
F\(1+y^{-2}\)^{-1}\cr =&{\rm const}(x-1)^{-k/N(1-k/N)}\widetilde
F\(1-x\)^{-1}~.}}
\fig{The same loops as in Fig 4.b after the loop $\bar\gamma_2$ has
been deformed around infinity.}{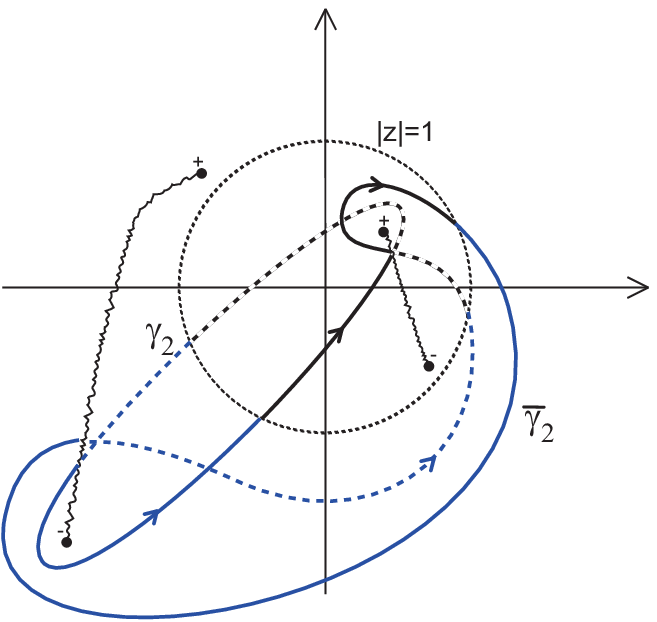}{8 truecm}
\figlabel\defgammatwo

\subsec{The classical piece}

If we are studying an orbifold of a torus, then there are also
classical contribution to the correlators, coming from worldsheet
instantons. These are solutions to $\d\bar\d X = 0$ which satisfy
the monodromy conditions \monodcondit, for the contour and values of
$v$ appropriate to the boundary conditions.

A general element of the combined orbifold group acts as
\eqn\Xtrans{X-X_0\ \to\ e^{2\pi ik/N}(X-X_0)+v~,}
where $v\in\Lambda$ is a point in the Narain lattice and $X_0$ is a
fixed point. In this subsection we will set $X_0=0$ and assume that the two twist operators
belong to the same fixed point. (The generalization to other fixed points, and to correlators of
twist operators belonging to different fixed points, are easily obtained from our results.)
Twist fields are built from local worldsheet insertions
about which $X$ has this monodromy.  Following \DixonQV,
we will label such insertions by $(\theta^k,v)$, where $\theta$ is
the generator of $Z_N$.

In order to construct a twist operator that is invariant under the orbifold
group, we have to sum over all of its images under conjugation by
elements of the orbifold group. Starting from some element
$(\theta^j,v_0)$ and conjugating it by $(\theta^k,u)$, we get
$(\theta^j,\theta^kv_0+(1-\theta)u)$ \DixonQV. Therefore, in the
orbifold theory, a twist operator is labeled by a {\it point group}
element $\theta^j$ and a {\it coset lattice} vector
\eqn\cosetlabel{\{\theta^kv_0+(1-\theta)u,\ k\in \IZ,\
u\in\Lambda\}~.}

As we encircle $|z| = 1$, the field $X$ transforms in a way dictated
by the twist operators inserted inside. Since the combined orbifold
group is non-commutative, that transformation depends on their
ordering. Here we choose a radial ordering centered at the origin.
Note that before summing over the coset lattice (and, by that,
obtaining a good operator in the orbifold theory), radial orderings
centered at different points (corresponding to different conformal
frames) are not equivalent.

In addition to the orbifold projection onto states
invariant under \Xtrans, the models we will consider
also contain orientifolds, so we must perform an orientifold
projection. An orientifold may wrap each of the $T^2$ factors
parametrized by $X$ ($\epsilon_1=1$) or be localized at one of the
fixed points $X_0$ ($\epsilon_1=-1$). A wrapped orientifold acts by
$\Omega$ and a localized orientifold acts by $R\Omega$, where
$\Omega$ is the worldsheet parity and $R$ is the reflection
\eqn\riflection{R:\qquad X-X_0~\longrightarrow~ -(X-X_0)~.}
In the case of $Z_3$ orbifolds of $T^2$, the fixed point $X_0$ is invariant where the
other two fixed points are identified. Again, here we set $X_0=0$.

\subsubsec{Monodromy and boundary conditions -- choosing contours}

When computing the quantum piece in section 2.3, we learned that for
$g_{-\pm}$ and $\epsilon_1=\pm 1$,
the non-trivial monodromy condition is obtained from
the loops labeled $\gamma_1$ in Figures 1-4,
while for $g_{-\pm}$ and $\epsilon_1=\mp 1$ it is
obtained from $\gamma_2$. One
difference between the quantum and the classical cases is that for
some correlators the two loops ($\gamma_1$ and $\gamma_2$) give
equivalent monodromy conditions. For the classical piece however,
one of the two conditions is always trivially satisfied.

To see this explicitly, note that for $g_{-+}$ with $\{\epsilon_1=1,\eps_2 = -1\}$,
and for $g_{--}$ with $\{\epsilon_1=-1,\eps_2 = 1\}$ the monodromy condition obtained by
translating the field $X$ around the loop $\gamma_2$ equals the
monodromy condition obtained by translating the field $X$ around the
loop $\gamma_1$, minus the monodromy condition obtained by
translating the field $\theta X$ around the loop $\gamma_1$. Since
the total change in the value of $X_{q}$ along the
closed loops was zero, the two loops $\gamma_1$ and $\gamma_2$
gave equivalent monodromy conditions, related by a non-zero number
($1-\theta$) \gpintm. However, the total
change in the value of $X_{cl}$ along the closed loops is some coset
lattice vector. That coset lattice vector is dictated by the twist
operators enclosed by the loop {\it and} their radial
ordering.\foot{Since the point group is commutative, for $X_q$ the
ordering of the group elements was not an issue.} Let the twist
operators encircled by $\gamma_1$ be $(\theta^i,v)$ at $y$ and
$(\theta^{-i},u)$ at the origin. As we translate $X_{cl}$ around
$\gamma_1$, it shifts by $v+\theta^iu$. The translation of $X_{cl}$
around $\gamma_2$ is equivalent to the translation of $X_{cl}$ around
$\gamma_1$, minus the translation of $\theta^i\widetilde X_{cl}$
around $\gamma_1$, where $\widetilde X_{cl}$ is the classical
solution with the ordering of the two twists reversed. Now since
$\theta^i(u+\theta^{-i}v)=v+\theta^iu$, the two shifts (of $X_{cl}$
and $\theta^i\widetilde X_{cl}$) cancel. Therefore, in these
specific cases, the monodromy condition obtained by translating the
field $X_{cl}$ along the loop $\gamma_2$ is trivially satisfied with $v = 0$.
Similarly for other boundary conditions and twist insertions,
the monodromy condition for one contour ($\gamma_1$ or $\gamma_2$)
is always trivially satisfied with $v = 0$. Therefore, the classical
solutions are labeled by a single vector $v$; one then performs the
sum in \fullcorrelator\ over the elements $v$ of the appropriate
coset lattice of the Narain lattice.

\subsubsec{Monodromies and boundary conditions -- the coset lattice}

For the disk or $RP^2$, the classical solution has to satisfy \discretesym\ at $|z|=1$.
This potentially affects the coset lattice which defines a good twist field in the conformal
field theory.

For $\epsilon_1=1$, \discretesym\ is automatically satisfied.  There
is no constraint on the possible twist insertions inside the
fundamental domain $|z|\leq 1$ of \discretesym.

For $\epsilon_1=-1$, the twist field insertions are restricted by
the demand that there is no non-trivial monodromy on the boundary.
That can be seen by noting that after moving the loop through the
boundary, it returns with the opposite orientation, so that the
corresponding monodromy must vanish.

Now, suppose that for $\eps_1 = -1$, we insert two twist fields,
labeled by $(\theta^i,v)$ at the origin and $(\theta^j,u)$ at $0<y<
1$. Consistency with the conditions at $|z|=1$ requires that
$u=-\theta^jv$.\foot{Since we put the brane at a fixed point $X_0$,
$X_0$ can "transform" to $\theta^{i+j}X_0=X_0$, so $i$ can be
different from $-j$.}\ \foot{Note that for the projective plane, if
$i\ne -j$ then $X=X_0$ at $|z|=1$.} Therefore, while for $\eps_1 =
1$ we have two independent sums over $u$ and $v$ in the coset
lattice for each of the two twist operators, for $\eps_1 = -1$ we
have only one sum over $v$, with $u$ constrained to be
$u=-\theta^jv$.

In the cases $\eps_1 = -1$, the nontrivial monodromy conditions come
from contours surrounding a twist field and its image under the map
\modding.  The image of $(\theta,v)$ is
$(\theta^{-1},-\epsilon_1\theta^{-1}v)$. Therefore, the monodromy
integral for a loop that encircles a twist operator $(\theta,v)$ and
its image $(\theta^{-1},-\epsilon_1\theta^{-1}v)$ on the sphere, is
equal to $(\theta^{-1}-\epsilon_1\theta^{-1})v$. Such a loop will be
non-trivial only for $\epsilon_1=-1$. In that case the coset lattice
is rescaled by 2 -- in other words, for $\eps_1 = -1$, we sum over
even multiples of the vectors $v$ which contribute for $\eps_1 = 1$.

\subsubsec{The full instanton sum}

By an explicit calculation we find that
\eqn\classical{H^{\eps_1,\eps_2}_{cl;-\pm}=\sum_v \exp\(-{\pi
R^2(5-3\epsilon_1)(5-3\epsilon_1\epsilon_2)\over 16\alpha'\sin(\pi
k/N)}\[{\widetilde F(1-x)\over \widetilde
F(x)}\]^{\epsilon_1}|v|^2\)~,}
where all the dependence on the sign $\pm$ of the twist at
$z_2,\bar{z}_2$ and some of the dependence on $\epsilon_2$ is hidden
the definition of $x$, as given in \xxx. The sum in \classical\ is
over the whole coset lattice; the rescaling by $2$ for $\eps_1 = -1$
is encoded in the prefactor $(5-3\epsilon_1)/2$ and the prefactor
$(5-3\epsilon_1\epsilon_2)/2=k^2$ (where $k^2$ is defined in \monodromy).
We have also
rescaled $v \to 2\pi R v$; that is, the cycles of the torus are
taken to have physical length $2\pi R$, but we define the generators
of $\Lambda$ to have length $1$.

\subsubsec{Relation to Dixon et. al.}

To see how this relates to the calculation of the four point
function on the sphere \DixonQV, recall that in that work the
monodromy condition for the classical piece $X_{cl}$ is
$$\Delta_{\LL_i} X_{cl}=\oint_{\LL_i} dz\d X_{cl}+\oint_{\LL_i} d\bar
z\bar\d X_{qu}=v_i~,$$
where $i\in\{1,2\}$, the loop $\LL_1$ ($\LL_2$) encloses
the points $w_2$ and $w_1$ ($w_2$ and $w_3$) as defined in \DixonQV,
and $v_i$ runs over a coset lattice which depends on which twist fields are
enclosed by $\LL_i$.

If we set
\eqn\fourtotwomp{
    (w_1,w_2,w_3,w_4) = (z_1,z_2, \epsilon_2/\bar{z}_2,\epsilon_2/\bar{z}_1)
}
for the case of $H_{-+}$, and
\eqn\fourtotwomp{
    (w_1,w_2,w_3,w_4) = (z_1,\epsilon_2/z_2, z_2,\epsilon_2/\bar{z}_1)
}
for the case of $H_{--}$, then we have just learned that in our case
$$\Delta_{\LL_i} X=0~,$$
where in the notations of \DixonQV\ we use in this section
$i=(3+\epsilon_1)/2$.

Once we have the classical solution, we must remember to integrate
it over $|z|\leq 1$ and not over the entire
complex plane. We conclude
that, if we define $x$ as in \xxx\ (see equation (4.37)), $S_{cl}$
is $\half{5-3\epsilon_1\over 2}{5-3\epsilon_1\epsilon_2\over 2}$
times the one in \DixonQV, where for $\epsilon_1=1$ $v_2=0$,
$\epsilon_1=-1$ for $v_1=0$ and when the argument of $F$ is bigger
than $1$, we replace it with $\tilde F$ (see Appendix $A$).

Alternately, one can check that the most general classical solution given in \DixonQV\ respects
\zbarzrel\ only when $v_2=0$ for $\epsilon_1=1$ and $v_1=0$ for
$\epsilon_1=-1$.

\newsec{Fermionic correlation functions}

Twist fields for fermions can be constructed via bosonization, and
{\it can}\ be factorized into holomorphic and antiholomorphic parts.
Thus the correlators are simply computed via the doubling trick,
without need of the conformal Ward identities.

Let us first review the twist fields and the bosonization map.
The twist fields $s_{\pm}$ are defined by the OPEs:
\eqn\twistopes{
    \eqalign{
    \psi(z)s_+(w)\sim &(z-w)^{k/N}t'_+(w)\cr
    \psi^*(z)s_+(w)\sim &(z-w)^{-k/N}t_+(w)\cr
    \bar\psi(\bar z)\bar s_+(\bar w)\sim &(\bar z-\bar w)^{-k/N}\bar t_+(\bar w)\cr
    \bar\psi^*(\bar z)\bar s_+(\bar w)\sim &(\bar z-\bar w)^{k/N}\bar
    t'_+(\bar w)~.
}}
The explicit construction of $s_{\pm}$ shows that it can be so
factorized into holomorphic and antiholomorphic pieces. Begin with
the bosonization formula for $\psi$:\foot{Here we suppress the
cocycles, as they will not be relevant for our computation.}
\eqn\bosonizationmap{
\eqalign{
    \psi(z) & = e^{i H(z)}\ ; \psi^*(z) = e^{-i H(z)}\cr
    \bar{\psi}(\bz) & = e^{i \bar{H}(\bz)}\ ; \bar{\psi}^*(\bz)
        =   e^{-i \bar{H}(\bz)}~.}}
With this definition, the twist fields are simply defined:
\eqn\fermtwistdef{
    s_{\pm}(z) = e^{\pm i k H(z)/N}\ ;\ \ \bar{s}_{\pm}(\bz) = e^{\mp i k
    \bar{H}(\bz)/N}~.}
Given this definition, the excited fermionic twist fields $t_{\pm},t'_{\pm}$, and so on, are
simple to construct as well.
Note that if $\psi$ has left-moving $R$ charge $1$, $s_{\pm}$ has fractional left-moving
R charge $\pm k/N$.

To describe the doubling trick for the orbifolds we study, we need the
action of \discretesym\ on the fermionic superpartners of $X$, and on the twist fields.
In the case of $\epsilon_1 = -1$, worldsheet supersymmetry requires that
we combine the action of \discretesym\ with a $Z_2$ subgroup of the $U(1)_R$
symmetry of the model. The resulting action of \discretesym\ on $\psi$ is:
\eqn\zbarzrel{\psi(z)=\epsilon_1 \bar\psi(\bar u)\({\d \bar u\over\d
z}\)^\half\|_{\bar u=\epsilon_2/z}=\epsilon_1
\bar\psi(\epsilon_2/z)\({-\epsilon_2\over z^2}\)^\half}
for $|z|>1$. The action of \discretesym\ on the twist fields must,
by consistency, be:
\eqn\twisttransf{
\eqalign{
    s_{\pm}(z) & = e^{\pm i(1 - \eps_1)\pi k/2N}\bar{s}_{\mp}(\bar{u})
    \left(\frac{\d \bar{u}}{\d z}\right)^{\frac{k^2}{2N^2}}\biggl|_{\bar{u} = \frac{\eps_2}{z}}\cr
    \bar{s}_{\pm}(\bz) & = e^{\mp i(1 - \eps_1)\pi k/2N}    s_{\mp}(u)
    \left(\frac{\d \bar{u}}{\d z}\right)^{\frac{k^2}{2N^2}}\biggl|_{u =
    \frac{\eps_2}{\bz}}~,}}
where $\Delta = \frac{k^2}{2N^2}$ is the holomorphic (antiholomorphic)
dimension of $s_{\pm}$ ($\bar{s}_{\pm}$).

The relevant correlators for \S4\ are, at this point, simple to calculate:
\eqn\ftwistcorr{
    \langle s_{-}(z_1)\bar{s}_-(\bz_1)s_{\pm}(z_2)
    \bar{s}_{\pm}(\bz_2) \rangle\biggl|_{z_1 = 0,z_2 = y\in \IR}
    = e^{\mp (1 - \eps_1) i \pi k/N}
    |y|^{\mp 2 k^2/N^2} |1 - \eps_2 y^2|^{-k^2/N^2}~.}

\newsec{Twisted sector kinetic terms to $\CO(g_s)$ in type I on $T^6/\IZ_3$}

In this section we will describe an application of our results
to the calculation of $\OO(g_s)$ (disk and $RP^2$) corrections to
the kinetic terms of certain twisted sector blow-up modes, for the
$T^6/\IZ_3$ compactification of type I string theory described in
\refs{\AngelantonjUY}. In \S4.1\ we will describe the orbifold and
the effective 4-dimensional picture for our calculation. In \S4.2\
we will construct the blow-up mode vertex operator. In \S4.3\ we
will compute the various pieces of the amplitude coming from the
disk and $RP^2$ with twist-twist and twist-anti twist insertions. In
\S4.4\ we will join the various pieces together by closely studying
the spacetime physics of the boundaries of the moduli space of  CFT
correlators.
In \S4.5\ we will write the full amplitude for the type
I orbifold, and further discuss the spacetime physics.

\subsec{The $T^6/\IZ_3$ orbifold and the 4d effective action}

We consider the type I oribifold on $T^6/\IZ_3$, as described by
\refs{\AngelantonjUY,\PoppitzDJ}.
We parameterize the six-torus $T^6$ by three complex coordinates
$X_{1,2,3}$, subject to the periodicity conditions
\eqn\tsixperiodicity{X_j \sim X_j + 2\pi R \sim X_j + 2\pi R \alpha~,}
where $\alpha = e^{2\pi i/3}$. This torus has a $\IZ_3$ symmetry,
under which the complex coordinates $X_j$ transformed as
\eqn\zthreetrans{(X_1, X_2, X_3) \rightarrow (\alpha X_1, \alpha
X_2, \alpha X_3)~.}
The transformation \zthreetrans\ has 27 fixed points, located at
$X_i = m_i\(2\pi R e^{i\pi/6}\)/\sqrt{3}$, with $m_i = 0,1,2$. The
resulting orbifold is a singular limit of a Calabi-Yau manifold with
Euler character $\xi=72$. The type II and heterotic versions of this
orbifold were constructed in \DixonJW\ and its geometry was analyzed
in detail in \StromingerKU. Each of the 27 fixed points has a
blow-up mode which leads to a modulus of the 4-dimensional effective
field theory at closed string tree level, and which resolves the
orbifold singularity; that remains true for the type I theory. The
vertex operators for these blow-up modes are in the twisted sector
of the corresponding fixed points. We may consider $C_k = \zeta_k +
i c_k$ to be a closed string chiral multiplet. The real part
$\zeta_k$, is the NS-NS twist field which blows up the orbifold; in
the type I orbifold, the imaginary part $c_k$ is a Ramond-Ramond
axion.

This theory has no D5-branes; the orbifold projects the type I gauge
group down to $U(12)\times SO(8)$. The $U(1)$ factor is anomalous;
the anomaly is cancelled by a 4d version of the Green-Schwarz
mechanism. This requires that the Ramond-Ramond axion $c$ in the
symmetric combination $C = \frac{1}{\sqrt{27}} \sum_k C_k=\zeta + i
c$ transforms by a shift under the open string $U(1)$. The result is
that the K\"ahler potential of the low-energy theory can be written
as:
\eqn\Kahlerpot{
    {\cal K} = {\cal K}(C+\bar C- g_sV)
}
where $V$ is the vector superfield for the anomalous $U(1)$. Note
the relative factor of $g_s = \langle e^{\phi} \rangle$ in the
gauging of the shift symmetry of $c$ (here $\phi$ is the dilaton).
This ensures that the tree-level couplings of the form
$\frac{c}{g_s}\tr F\wedge F$ cancel the anomalous transformation of
the path integral at one loop.

The original motivation for this work was to study the one-loop correction
to the Fayet-Iliopoulos D-term for this $U(1)$.  We will discuss this further
in \refs{\LS}; here we note that in many string models this controls
the tension of interesting cosmic D-strings \refs{\DvaliZH,\BinetruyHH},
and controls the decay constant of some interesting candidates for QCD
axions \refs{\SvrcekYI}.

The one-loop correction arises from
an $\CO(g_s)$ correction to the K\"ahler potential for $X$. To see this, note that
to one-loop order in $g_s$ and to
quadratic order in $C$ the $d^4\theta$ terms in the $N=1$ theory should have the form:
\eqn\Kehler{K=\frac{1}{g_s^2}(1+bg_s+dg_s^2)(C+\bar C-g_s
V)^2+\frac{1}{g_s}(a+cg_s)V+ \ldots~.}
Here, $a,b,c,d$ are numerical coefficients, and we have
suppressed the fluctuations of the dilaton. Explicit calculations
\refs{\DouglasSW,\PoppitzDJ}\ have shown that $a = c = 0$.
In other words, to one loop order, the Fayet-Iliopoulos D-term
vanishes at the orientifold point. The tree-level kinetic term for $C$ induces
a field-dependent FI D-term at open string tree level \refs{\DouglasSW},
which is nonzero for the blown-up orbifold.

It had been conjectured in \refs{\PoppitzDJ,\LawrenceSM}\ (and stated
in various works since) that the one-loop correction to the field-dependent
D-terms would vanish away from the orientifold point.
In particular this would imply that $b=0$.  However, from
an effective field theory standpoint, we see no reason for $b$ to vanish.  In the remainder
of this section we will calculate this term and find that it is in fact non-vanishing.

We will extract $b$ from the $\CO(k^2)$ part of the two-point
functions of twist fields on the disk and $RP^2$. For truly massless
fields the on-shell coondition is $k^2=0$.
However, the FI D-term
above gives a positive $\CO(g_s)$ mass to a linear combination of
$C$ and the charged scalars. More precisely, the kinetic term for
the anomalous $U(1)$ gauge field is:
\eqn\anabelkin{
    L = V_{6}\int d^2 \theta \left(\frac{\alpha}{g_s} + \Delta\right) \WW_{\alpha}\WW^{\alpha}
        + \ldots
}
where $\WW_{\alpha}$ is the chiral Fermi superfield for the gauge
multiplet, $g_{YM,tree} = \sqrt{g_s/\alpha}$ is the tree-level gauge
coupling (which will also generically depend on $C$), $\Delta$ is
the one-loop threshold correction to the gauge coupling and $V_6$ is
the volume of $T^6/\IZ_3$ in string units. If we integrate out the
auxiliary $D$ field in the gauge multiplet, the quadratic action for
$\zeta$ becomes:
\eqn\leea{I\sim{1\over g_s^2}\int
dx^4\[\d_\mu\zeta\d^\mu\zeta\(1+bg_s + d g_s^2\)- {g_s\over V_6} \(1
+ (2 b - \Delta) g_s  +\ldots\) \zeta^2\]~,}

String theory amplitudes compute amputated on-shell matrix elements.
The two-point function thus vanishes on shell, as it is proportional to the inverse
propagator. In the present case, the closed string field $C$ gets a mass from
disk and $RP^1$ diagrams. Thus the contribution of any given loop order will not vanish;
rather, the full two-point function vanishes through cancellations between different
loop orders. In particular the on-shell condition will be
$k^2 \sim g_s + ...$.  We will therefore compute $b$ by computing the $\CO(k^2)$ term for
the two-point function on the disk and $RP^2$.

\subsec{The closed string vertex operator}

The first step is to construct the vertex operators corresponding to
the blow-up mode of that fixed point. Let $\XX^{\mu=0,1,2,3}$
parameterize the non-compact dimensions (4d Minkowski space)
transverse to the orbifold; $c,\bar{c}$ be the anticommuting
conformal ghosts and $\phi,\ \bar\phi$ arise from the bosonization
of the superconformal ghosts (see for example
\refs{\FriedanGE,\PolchinskiRQ}). The vertex operator for the
massless blow-up mode in the $(-1,-1)$ picture is
\eqn\typeIItwistvo{V_{\pm1/3}^{(-1,-1)}(z,\bar
z;k)=\frac{1}{\sqrt{27}}e^{-\phi(z)-\bar\phi(\bar z)}e^{ik_\mu
\XX^\mu}\prod_{i=1}^3\(\sum_{p=0}^2 \sigma_{p,\pm}^{(i)}(z,\bar
z)s_\pm^{(i)}(z)\bar s_\pm^{(i)}(\bar z)\)~,}
where $k^2=0$. The sum over $p$ is a sum over the fixed points of
$T^2/\IZ_3$: since fixed points of $T^6/\IZ_3$ are in particular
fixed points of $\(T^2/\IZ_3\)^3$, the twist operators in
\typeIItwistvo\ are the product of the corresponding ones in each
$T^2/\IZ_3$ factor.

The orientifold action maps twisted sectors to their anti-twist
sectors, breaking the $\IZ_3$ quantum symmetry of the type IIB
orbifold \DineDK\ that would describe the closed string sector
absent the D9 branes and O9 planes of type I string theory. The
orientifold action interchanges $V_{1/3}$ and $V_{-1/3}\equiv\Omega
V_{1/3}\Omega$. Therefore only the combination
\eqn\typeIItwistvo{V^{(-1,-1)}= V_{1/3}^{(-1,-1)}+\Omega
V_{1/3}^{(-1,-1)}\Omega\equiv V_{1/3}^{(-1,-1)}+V_{-1/3}^{(-1,-1)}}
survives the orientifold projection.

\subsec{Computing the two-point amplitudes}

The total $\phi$ ($\bar\phi$) charge on a worldsheet
$h$ holes plus crosscaps and $g$ handles is
$2g+h-2$. For the disk and $RP^2$, $g=0$ and $h=1$. So we will take
one vertex operator to be in the $(-1,-1)$ picture and the other in
the $(0,0)$ picture. The $SL(2,\IR)$ symmetry of these amplitudes
allows us to fix the position of the two vertex operators to
$(z_1,z_2)=(0,y)$, where $y\in [0,1]$.

The D-branes and orientifold planes in type I
string theory fill the 10d spacetime, and so corresponds to the case $\eps_1 = 1$ in \S2.

The amplitudes on the disk should include a trace over Chan-Paton
factors. For correlation functions on the disk, an element $g$ of
the orbifold group $\Gamma$ acts on the Chan-Paton matrices $t$ as
in \refs{\GimonRQ,\DouglasSW}, by conjugation: $t\to \gamma_h t
\gamma_g^{-1}$, where $\gamma_{g}$ is the image of a group
homomorphism of $\Gamma$ into the open string gauge group.
For specificity, the matrices $\gamma_{\frac{1}{3}}$ for the $\IZ_3$
twists are given in \PoppitzDJ.

In order to compute the $\CO(g_s)$ correction to the kinetic term for
blowup modes, we must compute the disk correlators
\eqn\amplitude{ \eqalign{ \CA_{\pm,\pm}^{\eps_2=1}= & \tr \left(
\gamma_{\pm \frac{1}{3}}\gamma_{\pm \frac{1}{3}}\right)\times\cr &\
\ \ \ \times\ \int_0^1dy\<\xi\(c\bar c\ e^\phi T_Fe^{\bar\phi}\bar
T_FV_{\pm \frac{1}{3}}^{(-1,-1)}(0,0;k)\)\((c-\bar c)V_{\pm
\frac{1}{3}}^{(-1,-1)}(y,y;-\tilde k)\)\>~.}}
and the $RP^2$ correlators which are identical but for the absence
of the Chan-Paton trace and for an additional normalization factor
which we will determine in \S4.4. Here $\xi$ is the zero mode in the
superconformal ghost sector as described in
\refs{\FriedanGE,\PolchinskiRQ}.

The fermionic stress tensor is
\eqn\fermst{
    T_F(z)=-\frac14\sum_{i=1}^3(\d X_i\psi_i^*+\d\bar
    X_i\psi_i)-\half\d\XX\cdot\Psi\ ,}
with a similar definition for $\bar T_F(\bar z)$. $e^{\phi}T_F V(z)$
denotes the simple pole in the OPE of $e^{\phi} T_F(w)V(z)$, and is
the result of acting with the picture changing operator on $V$. Note
that for the disk and $RP^2$, only the sum of left- and right-moving
ghost number is conserved. Therefore, we can equally well do the
computation with both vertex operators in the $(-1,0)$ picture, or
with one vertex operator in the $(-1,0)$ picture and the other in
the $(0,-1)$ picture.

We first compute
\eqn\minplus{\eqalign{
    N^{\eps_2}_{-+} &\int_0^1dy\<\xi \(c\bar c\ e^\phi
T_FV_{1/3}^{(-1,-1)}(0,0;k)\)\((c-\bar c)e^{\bar\phi}\bar
T_FV_{-1/3}^{(-1,-1)}(y,y;-\tilde k)\)\>\cr
    \equiv &\int_0^1dy \HH^{\eps_2}_{-+}(y)~.}}
where $N^{\epsilon_2}_{-\pm}$ is a normalization factor. Following
\DixonQV, we can simplify the amplitude by examining the action of
the picture-changing operators more closely. In the OPE of the
compact part of $T_F$ with $\sigma_+^{(i)}s_+^{(i)}$, only the $\d
X_i\psi_i^*$ term is singular. Similarly, in the OPE of the compact
part of $\bar T_F$ with $\sigma_-^{(i)}\bar s_-^{(i)}$, only the
$\bar\d X_i\bar\psi_i^*$ term is singular. The B-type boundary
conditions we study here conserve the sum of the holomorphic and
antiholomorphic fermion number \refs{\OoguriCK}. Therefore, this
part of the amplitude vanishes by fermion number conservation, and
only the non-compact part of $T_F$ ($\bar T_F$) involving
$-\half\d\XX\cdot\Psi$ ($-\half\bar\d\XX\cdot\bar\Psi$) contributes
to this amplitude. The nonvanishing part of $H_{-+}(y)$ is:
\eqn\nonvanishingpm{ \eqalign{\HH^{\eps_2}_{-+}(y)=&
N^{\eps_2}_{-+}\sum_{p,q=1}^{27}\<e^{-\bar\phi}(0)e^{-\phi}(y)\>\<c(0)\bar
c(0)[c(y)-\bar c(y)]\>\cr \times&\<e^{ik\cdot\XX}(0)e^{-i\tilde
k\cdot\XX}(y)\>\<k\cdot\Psi(0)(-\tilde k)\cdot\bar\Psi(y)\>\cr
\times&\<\sigma_{+,p}(0,0)\sigma_{-,q}(y,y)\>\<s_+(0)\bar s_+(0)s_-(y)\bar
s_-(y)\>~,}}
where $\sigma_{\pm,p}$ and $s_{\pm}$ are shorthands for the products
of $\sigma_{\pm, p}^{(i)}$ and $s_{\pm}^{(i)}$ over the three $T^2$
factors of $T^6$, and $p,q$ label the orbifold fixed points. Note
that the term $\<e^{ik\cdot\XX}(0)e^{-i\tilde k\cdot\XX}(y)\>
\<k\cdot\Psi(0)(-\tilde k)\cdot\bar\Psi(y)\>$ is proportional to
$k^2$.

The second term in \amplitude\ that we have to compute is
\eqn\pplus{\eqalign{ N^{\eps_2}_{--}&\int_0^1dy\<\(c\bar c\ e^\phi
T_F V_{-1/3}^{(-1,-1)}(0,0;k)\)\((c-\bar
c)e^{\phi}T_FV_{-1/3}^{(-1,-1)}(y,y;-\tilde k)\)\>\cr
\equiv&\int_0^1dy \HH^{\eps_2}_{--}(y)~,}}
and similarly for $H_{++}(y)$. As for $H_{-+}$, when choosing the
appropriate picture, only the 4d spacetime parts of $T_F$ contribute
to the amplitude. The result is
\eqn\nonvanishingpp{\eqalign{\HH^{\eps_2}_{--}(y)=
&N^{\eps_2}_{--}\<e^{-\phi}(0) e^{-\phi}(y)\>\<c(0)\bar
c(0)[c(y)-\bar c(y)]\>\cr \times&\<e^{ik\cdot\XX}(0)e^{-i\tilde
k\cdot\XX}(y)\>\<k\cdot\bar\Psi(0)(-\tilde k)\cdot\bar\Psi(y)\>\cr
\times&\<\sigma_-(0,0)\sigma_-(y,y)\>\<s_-(0)\bar s_-(0)s_-(y)\bar
s_-(y)\>~.}}

The results for $\HH_{+-}$ and $\HH_{++}$ will be identical to the
above, and we will not discuss them explicitly.

\subsubsec{The ghost and spacetime parts of the amplitudes}

The contributions of the ghosts and spacetime parts are:
\eqn\restofem{ \eqalign{ &\<e^{-\phi(0)}e^{-\phi(y)}\>= 1/y\cr
&\<e^{-\bar\phi}(0)e^{-\phi}(y)\>= \omega_{\phi}\cr
&\<k\cdot\Psi(0)(-\tilde k)\cdot\bar\Psi(y)\> = \omega_{\psi}
k\cdot\tilde k\cr &\<k\cdot\Psi(0)(-\tilde k)\cdot\Psi(y)\> =
k\cdot\tilde k/y\cr &\<c(0)\bar c(0)[c(y)-\bar c(y)]\> =\omega_c
y\cr &\<e^{ik\cdot\XX(0,0)}e^{-i\tilde k\cdot\XX(y,y)}\>=\omega_\XX
\delta^4(k-\tilde k)\[{1-\epsilon_2y^2\over y^2}\]^{k^2} ~.}}
Here $\omega_{\phi,\psi,c,\XX}$ are phase factors which we will set
later using spacetime considerations, $\delta^4(k-\tilde{k})$ comes
from integration over the zero modes in $\IR^4$ and is a volume
divergence.

\subsubsec{Twist field correlators and the coset lattice}

The correlation function of twist fields can be broken up following
\fullcorrelator\ into a classical and a quantum piece.  The quantum
part of the correlator is independent of the particular fixed point on which
the classical solution resides.




To find the right lattice, note that the two-point function of $\Phi$ breaks up into
a sum over fixed points.
\eqn\breakup{
    \langle \Phi\Phi \rangle = \frac{1}{27}\sum_{p,q}\langle \Phi_p \Phi_q\rangle
    = \sum_q \langle \Phi_p \Phi_q \rangle
}
where the last line is $p$-independent.
As shown in \S3\ of \refs{\DixonQV} (\cf\ Fig. 3 of that work),
the sum over $q$ and the coset lattice \cosetlabel\ is just the sum over the
full Narain lattice $\Lambda$. We can define this lattice as:
\eqn\lattice{\Lambda = \{\eta(n_1,n_2,n_3)+\bar\eta(m_1,m_2,m_3)\ |\
n_i,m_i\in\ZZ ,~\eta=\frac{1}{\sqrt{3}}(1+e^{i\pi/3})\}~.}
Once we have this, the results of \S2\ and \S3\ for $\eps_1 = 1$
give us the twist field contributions to the two-point function.

\subsubsec{The full amplitude}

The full amplitude is:
\eqn\finalpm{ \eqalign{\HH^{\eps_2}_{-\pm}(y)=
&N^{\eps_2}_{-\pm} \omega_{tot}{k^2\over y}{1\over 1-\epsilon_2
y^{\pm2}}
\[{1-\epsilon_2y^{\pm2}\over y^{\pm2}}\]^{k^2}\delta^4(k-\tilde k)\cr
\times&\[\widetilde
F^{-1}(\epsilon_2y^{\pm2})\sum_{m,n=-\infty}^\infty \exp\(-{\pi
R^2(5-3\epsilon_2)\over 12\sqrt{3}\alpha'}{\widetilde
F(1-\epsilon_2y^{\pm2})\over \widetilde
F(\epsilon_2y^{\pm2})}|n\eta+m\bar\eta|^2\)\]^3
~.}}
%
Here the phase $\omega_{tot}$ is the product of all of the phases
$\omega_{\phi,\psi,c,\XX}$.

We wish to add together all of the amplitudes
$\int_0^1dy\HH^{\eps_2}_{\pm,\pm}$. We will do so in the next sub-section using
tadpole cancelation arguments.

\subsec{Connecting the pieces}

Define
\eqn\transformtoline{\eqalign{&\widetilde
\HH^{\eps_2}_{-\pm}(x)\equiv {1\over\delta^4(k-\tilde
k)}\HH^{\eps_2}_{-\pm}(y){dy\over dx}\cr\cr
=&\NN_{-,\pm}^{\eps_2}{k^2\over x(1-x)}\[{1-x\over x}\]^{k^2}
\[\widetilde F^{-1}(x)\sum_{m,n=-\infty}^\infty \exp\(-{\pi
R^2(5-3\epsilon_2)\over 4\sqrt{3}\alpha'}{\widetilde F(1-x)\over
\widetilde F(x)}|n\eta+m\bar\eta|^2\)\]^3
~.}}
Here $\NN$ is the product of all numerical factors. The variable $x$
is defined in \xxx\ -- its range depends on the relative sign of the twist fields, and
on the boundary conditions, as shown in figure 6. The sole dependence
of the amplitude on the relative signs of the twist fields is encoded in the range
of $x$.

As a result, we can consider the full amplitude -- the sum of the
modular integrals of the amplitudes over $\eps_2$ and over the signs
of the twist fields -- to be a modular integral of a single
integrand over the entire real $x$-axis. The integrand is one of the
functions $\HH^{\eps_2}_{\pm,\pm}$ depending on the value of $x$.
Note that for fixed $\eps_2$ and sign of the twist fields,
$\HH^{\eps_2}_{\pm,\pm}$ are not necessarily finite at the
boundaries of the range in $x$ over which they are defined. However,
the type I model in $T^6/\IZ_3$ is known to be a consistent string
background.  Therefore, any divergences must either cancel in the
full amplitude, or correspond to IR divergences arising from the
exchange of physical massless modes. We will use these facts to fix
the relative values of $\NN^{\eps_2}_{\pm,\pm}$.
%
\fig{The different piece of the $x$-integral. $x\in[-\infty,-1]$
comes from the projective plane with two identical twist insertions,
$x\in[-1,0]$ comes from the projective plane with twist and
anti-twist insertions, $x\in[0,1]$ comes from the disk with twist
and anti-twist insertions, $x\in[1,\infty]$ comes from the disk with
two identical twist insertions.}{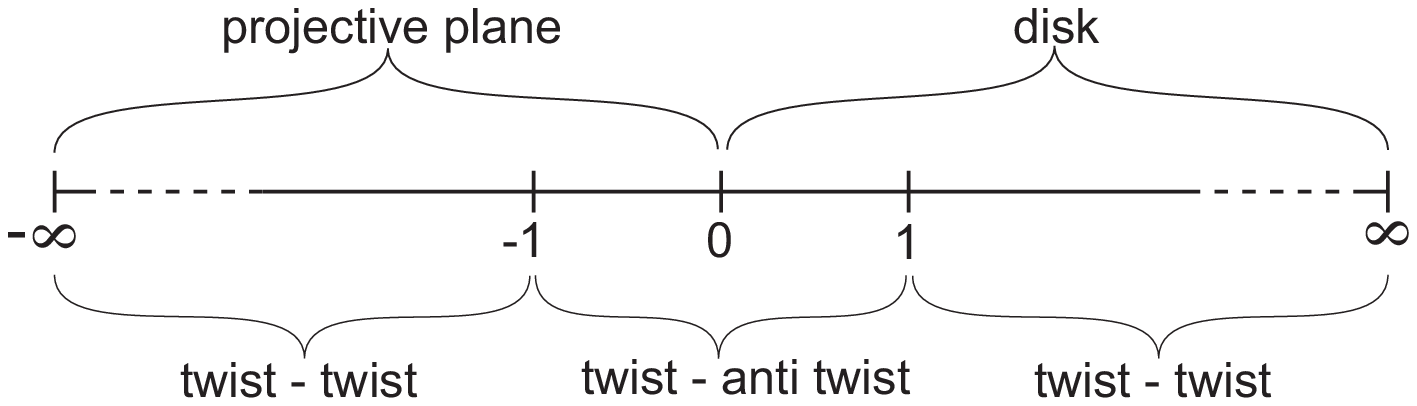}{12 truecm}
\figlabel\ehairpin

We begin by noting that the amplitudes for $\HH_{-\pm}$ and $\HH_{+\mp}$ will give
identical contributions since
%
\eqn\pmequivalence{\<V_{-1/3}V_{\pm 1/3}\>=\<V_{-1/3}\Omega V_{\mp
1/3}\Omega\>=\<\Omega V_{-1/3}\Omega V_{\mp 1/3}\>=\<V_{1/3}V_{\mp
1/3}\>~.}
Therefore, we need only determine four amplitudes, $\HH^{\eps_2 = \pm 1}_{-\pm}$.

Now let us discuss in turn the behavior near $x = -1, 0,\pm\infty,$ and $1$,
at which we must join together $\HH^{\eps_2}_{-\pm}$.

\subsubsection{The amplitudes near $x = -1$ -- continuity constraint}

The first relation between the coefficients $\NN^{-}_{-\pm}$ arises
from the fact that $x = -1$ is not actually a boundary of the moduli
space of the corresponding correlation function.
More precisely, the $RP^2$ amplitudes
$\widetilde\HH_{--}^{-}(-\infty\le x\le-1)$ and
$\widetilde\HH_{-+}^{-}(-1\ge x\le 0)$ connect continuously at
$x=-1$, as $V_{1/3}$ at $|z| = 1$ is equal to $\Omega
V_{1/3}\Omega=V_{-1/3}$ at $z = - 1/\bar{z}$. Therefore,
$\widetilde\HH_{--}^{-}(-1)dx$ and $\widetilde\HH_{-+}^{-}(-1)dx$
are the {\it same} amplitude, which implies that:
\eqn\xtominusone{{\NN_{--}^{-}\over \NN_{-+}^{-}}=1~,}

\subsubsection{The amplitude near $x = 0, \pm\infty$ -- tadpole cancellation constraints}

The remaining ponts $x = 0,\pm\infty,1$ are true boundaries of moduli space
of 2d correlators on the disk.
%
%
%
At $x\to 0,\infty$ the two closed string vertex operators approach
each other and the amplitude factorizes onto a three-point function
on the sphere times a tadpole diagram, joined by the propagator for
the massless closed string mode (see Figure 7).
In these limits the $x$ integral diverge.\foot{Remember that $k^2\sim
g_s>0$.} Since all tadpoles cancel for the type I model we are
studying \refs{\AngelantonjUY}, the disk and $RP^2$ contributions should be added such
that the divergences due to tadpoles cancel.

\fig{At one boundary of the moduli space, when the two closed string
vertex operators approach each other, the amplitude factorizes onto
a three-point function on the sphere times a tadpole diagram, here
drawn for the disk.}{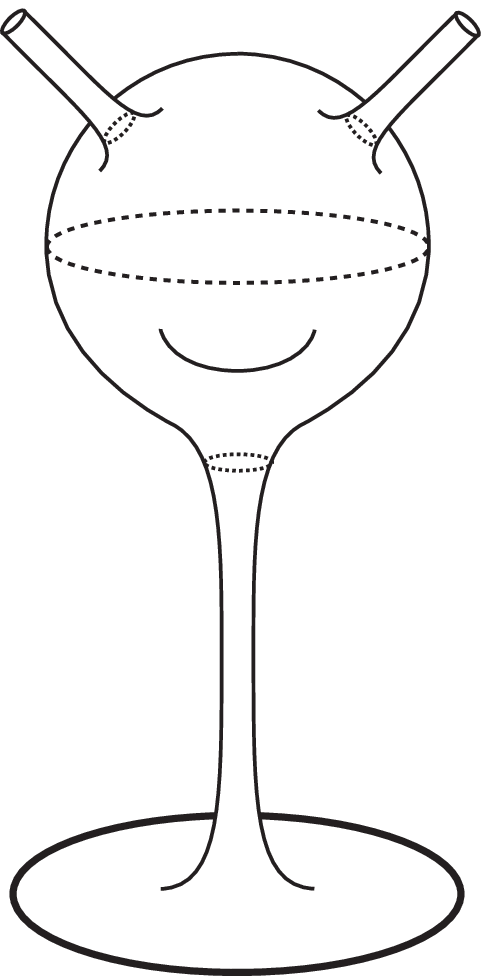}{4 truecm} \figlabel\ytozero

For $\HH^{\eps_2}_{--}$, the $RP^2$ tadpole at $x\to -\infty$ and
the disk tadpole at $x\to\infty$ should cancel.\foot{The
nonvanishing $(--)$ amplitudes at $\CO(g_s)$ follow from the
breaking of the quantum symmetry of the oriented closed string
theory. The orientifold already breaks the quantum symmetry, as the
orientifold action maps twisted sectors to their conjugates \DineDK.
Furthermore, the D-branes can also break this quantum symmetry, as
can be seen from the existence of twisted sector tadpoles that must
be cancelled with the correct choice of gauge group
\refs{\GimonRQ,\AngelantonjUY}.} This requires:
\eqn\xtoinfty{{\NN_{--}^{-}\over \NN_{--}^{+}}=-8~,}
where we have used the asymptotic behaviors:
\eqn\xtoinfhyper{
    F(-x)\sim{\Gamma(1/3)\over\Gamma^2(2/3)}x^{-1/3}~,\qquad \widetilde F(x)\sim
    \half{\Gamma(1/3)\over \Gamma^2(2/3)}x^{-1/3}~,\qquad\qquad x\to\infty~.}

For $\HH^{\eps_2}_{-+}$, the divergence in the disk tadpole as $x\to 0^+$ should cancel
the divergence in the $RP^2$ tadpole as $x\to 0^-$. This requires that
\eqn\xtozero{{\NN_{-+}^{-}\over \NN_{-+}^{+}}=1~,}
where we have used the asymptotic behaviors:
\eqn\xtoonehyper{
    F(x)\sim 1~,\qquad F(1-|x|)\sim \widetilde F(1+|x|)
    \sim -{\sin(\pi k/N)\over \pi}\ln(x)~,\qquad\qquad x\sim 0~.}

\subsubsection{The amplitude near $x = 1$ and open string dynamics}

The final boundary of moduli space is at $x = 1$. The amplitudes
$\HH^{+}_{-\pm}$ factorize in the limit $x\to 1^{\mp}$ onto two
closed string one-point function on the disk, joined by an open
string propagator (see Figure 8).  The relative weights of these
amplitudes are completely fixed by \xtominusone, \xtoinfty\ and
\xtozero\ to be:
\eqn\xtoone{{\NN_{-+}^{+}\over \NN_{--}^{+}}=-8~,}
We would like to check that this is consistent.

\fig{At one boundary of moduli space, when one of the closed string
operators approach the boundary of the disk, the disk amplitude
factorize into two disks with closed string twist insertions,
connected by a massless open string propagator.}{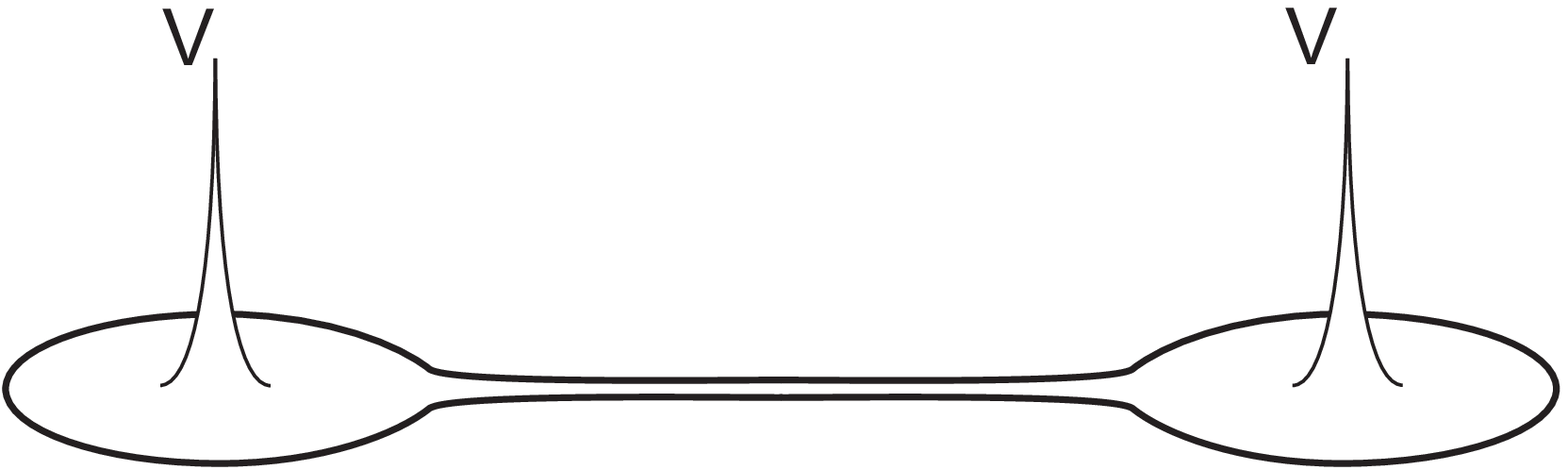}{12
truecm} \figlabel\ytoone

First, we note that \xtoone\ is completely consistent with the
Chan-Paton factors.  By doing Poisson resummations of the sums in
\transformtoline, using the formula
\eqn\Poisson{\beta\sum_{n,m}e^{-{\pi\beta\over\tau_2}|m-\tau
n|^2}=\sum_{k,w}e^{-{\pi\over\beta\tau_2}|k-\tau w|^2}~,}
we can see that $\HH^{+}_{--}$ and $\HH_{-+}^{+}$, have the same
behavior near $x=1$, with their normalization differing only by the
Chan-Paton factors.
 These are $\Tr(\gamma_-\gamma_-)$ and
$\Tr(\gamma_-\gamma_+)$, respectively.
Using the explicit expressions in \PoppitzDJ\ for $\gamma_{\pm}$:
\eqn\orbmat{\gamma_{\pm\frac13}= \left( \matrix{{\bf 1}_8 & 0 & 0\cr
0 & -\half {\bf 1}_{12} & \pm \frac{\sqrt{3}}{2} {\bf 1}_{12}\cr 0 &
\mp\frac{\sqrt{3}}{2} {\bf 1}_{12} &  -\half {\bf 1}_{12}}\right)~,}
where the ${\bf 1}_{d}$ is the $d\times d$ identity matrix,
we find that:
\eqn\agreement{{\Tr(\gamma_-\gamma_+)\over
\Tr(\gamma_-\gamma_-)}=-8}
in agreement with \xtoone.


The explicit behavior of the amplitude near $x = 1^{\pm}$ is:
\eqn\openfactordiv{\widetilde{\HH}_{-,\mp}
\sim\frac{k^2/R^6}{(1-x)^{1 + k^2}} ~,}
where the $R^6$ comes from the Poisson resummation. Thus, although
the amplitude appears to be proportional to $k^2$,the formula
\eqn\deltalimit{\lim_{k^2\to 0}\frac{k^2}{(1-x)^{1 +
k^2}}=\delta(1-x)~,}
shows that
there is a leading term which is constant as $k^2 \to 0$: it is an
$\CO(g_s)$ mass term. The $1/R^6$ in \openfactordiv\ matches the
$1/V_6$ factor in \leea.

Such a term could arise from integrating out the massless (at
leading order in $g_s$) $U(1)$ gauge field if it has a derivative
coupling to the physical blow-up mode (the mode that survives the
orientifold projection).
However, this coupling vanishes. The disk amplitude for this
open-closed coupling is
\eqn\twistgauge{
    A_{\pm} = \langle V_{\pm\frac{1}{3}}(0,0;k) \d \XX^{\mu}(1)e^{-ik\cdot \XX}(1)\zeta_{\mu}\rangle
    \left(\tr \gamma_{\pm\frac{1}{3}} t_{U(1)}\right) \propto k\cdot \zeta \tr
    \left(\gamma_{\pm\frac{1}{3}} t_{U(1)} \right)~,
}
where $\zeta_\mu$ is the polarization vector and $t_{U(1)}$ is the
$U(1)$ Chan-Paton matrix. The only difference
between $A_+$ and $A_-$ is the Chan-Paton trace.
Using $\gamma_{\pm \frac{1}{3}}$ \orbmat\ and $t$ as given in \PoppitzDJ:
\eqn\tmat{t_{U(1)}= \left(\matrix{0 & 0 & 0\cr 0 & 0 & {\bf
1}_{12}\cr 0 & -{\bf 1}_{12} & 0}\right)~,}
we find that
\eqn\tracecancel{
    \tr \left(\gamma_{\frac{1}{3}} t_{U(1)}\right)
     = - \tr \left(\gamma_{- \frac{1}{3}} t_{U(1)}\right)~.}
Thus \typeIItwistvo\ does not couple to any propagating massless
fields.

In fact, the origin of this term is as a contact term in the CFT.
Such contact terms have been discussed in the context of closed
strings in \refs{\AtickGY,\DineGJ}.  The interpretation here is the
same -- the contact term arises from integrating out the
non-propagating auxiliary field in the $U(1)$ $\NN=1$ supermultipet.
In this case, the open string channel can be thought of as
corresponding to the auxiliary field, and the disk diagrams
corresponds to tree-level dependence of the Fayet-Iliopoulos D-term
on the twist fields, which is known to be nonvanishing, as is clear
from \Kehler \refs{\AngelantonjUY,\PoppitzDJ}.\foot{See also
\BainFB.}

As a check that the disk diagrams are nonvanishing, in light of our
discussion of the vanishing of the scalar-gauge-field couplings above,
note that the corresponding vertex operator is
the $U(1)_R$ current of the model on the boundary
\refs{\AtickGY,\DineGJ,\LawrenceSM}. Since $V_{\pm\frac{1}{3}}$
have opposite worldsheet R-charge, the signs from the Chan-Paton
traces \tracecancel\ are cancelled by the signs from the R-charge sector.  The Chan-Paton
factors multiply a nonvanishing CFT amplitude.

%


\subsec{The full amplitude on $T^3/Z_3$}

As discussed in \S4.1, we are
expecting two distinct contributions to \leea\ from the disk and $RP^2$ diagrams.
One is the $\OO(g_s)$
mass for $\zeta$ (from integrating out the auxiliary D-field) and the other is
the $-g_sk^2b$ correction to the kinetic term for $\zeta$. The on-shell
condition is $k^2\sim g_s$ and our calculation should apply for any
(small) value of $g_s$. Therefore the distinction between the two
contributions to \leea\ are in the $k^2$ dependence of the disk and $RP^2$ diagrams,
before we put $k^2$ on-shell.

To extract the $\OO(g_s)$ mass term we put $k^2=0$.
The $\OO(g_s)$ mass term is reproduced in our amplitude from the
contact term that arises as $k^2 \to 0$, in the region of integration near $x=1$.
It comes with the numerical coefficient
\eqn\Dmass{c=\half\(\NN_{--}^{+}+\NN_{++}^{+}\)+
\half \(\NN_{-+}^{+}+\NN_{+-}^{+}\)=-7\NN_{--}^{+}~.}

To extract the correction to the kinetic term ($-g_sk^2b$), we hold
$k^2$ fixed, subtract the $\OO(g_s)$ mass term \Dmass\ and divide
the result by $k^2$. In order to extract the value of $b$ we then
take the limit $k^2\to 0$ -- only the contact term \deltalimit\ made
this order of limits
problematic, and we have subtracted it out.

More explicitly
\eqn\sumint{\eqalign{b_n\propto\lim_{k^2\to 0}{1\over k^2}
\[\int_1^\infty \widetilde \HH_{--}^{+}(x)dx + \int_0^1 \widetilde
\HH_{-+}^{+}(x)dx + \int_{-\infty}^0 \widetilde
\HH_{-+}^{-}(x)dx-c\]\equiv\tilde b~.}}

We do not know how to evaluate the integrals in \sumint\
analytically. Note that \sumint\ does not possess any obvious
symmetry property and therefore, in appose to \Dmass, we expect it
to depend on $R$ (the volume of $T^6/Z_3$). Using Mathematica we
evaluated it numerically in the non-compact limit ($R\to\infty$),
where the classical contribution \classical\ become trivial.\foot{It
is hard to maintain numerical precision in evaluating the
hypergeometric function $F(x)$ near $x=1$. To deal with this
problem, we use its explicit asymptotic behavior \xtoonehyper\ near $x = 1$, as
explained in \KlebanovMY.} We found the non-zero value for
\sumint:\foot{The poor accuracy arises because we are canceling
logarithmic divergences.}
\eqn\bval{\tilde b(R\to\infty)~= ~4.7 \pm .1~.}
Since we expect $b(R)$ to have a smooth limit as $R\to\infty$, we
take \bval\ as evidence that $b(R)\ne 0$ for generic $R$.

The FI D-term enters into various physical quantities such as the physical
mass of $\xi$ as shown in \leea.  Even though $b$ is nonvanishing,
one might wonder if $2b - \Delta$ in (4,6) vanishes at the orbifold point -- indeed, in the
type IIA model in \refs{\AkerblomUC}, it was argued that the one-loop correction
to the D-term is equal to the threshhold correction at the orientifold point.
In the present model this cannot be the case.  In particular, the instanton
corrections give $b$ an explicit dependence on the untwisted
K\"ahler moduli $R$. However, at the orbifold point
the cylinder and M\"obius strip amplitudes at constant field
strength \refs{\AntoniadisGE}\ from which one extracts the threshold correction $\Delta$
are independent of the untwisted K\"ahler moduli.\foot{The one-loop correction to the
gauge coupling anomalous $U(1)$ is not computed in \refs{\AntoniadisGE}, as it
requires an additional subtraction of divergences.  Nonetheless, the cylinder and M\"obius
strip diagrams which must contribute to the answer {\it are}\ computed there, and
they are independent of the untwisted K\"ahler moduli.}
Thus, we expect that for general $R$, neither the one-loop correction to the
field-dependent FI D-term, nor the physical mass for $\xi$, will vanish.



\newsec{Conclusions}

Since \bval\ is nonvanishing, \Kahlerpot\ indicates that in type I
compactification on $T^6/Z_3$ there will be a correction to the
Fayet-Iliopoulos D-term at order $g_s$ and to first order in a
perturbation of the theory away from the orientifold point. This is
consistent with the explicit calculation in \PoppitzDJ\ which shows
that the one-loop correction vanishes {\it at}\ the orientifold
point. It contradicts the conjecture that this correction would
vanish away from the orientifold point given in
\refs{\PoppitzDJ,\LawrenceSM}, but there were no strong arguments
for this conjecture, and no clear reason for it to hold from the
standpoint of 4d effective field theory.

It is sometimes stated that the results of \FischlerZK\ imply that
there should be no contributions at higher loops. However, these
results depend on the field theory being renormalizable and on the
gauge symmetry in the UV being linearly realized.  These conditions
are absent in our string model and in many others, and we see no
reason for the nonrenormalization theorem to hold -- in particular,
the coefficient $d$ in \Kehler\ leads to a field-dependent FI D-term
at two loops in open string perturbation theory. From the standpoint
of effective field theory, we see no reason for this term to vanish,
and suspect that it does not.  We will discuss the effective field
theory point of view in more detail in \LS.

\bigskip
\centerline{\bf Acknowledgements}

We would like to thank Allan Adams, Howard Schnitzer and Eva
Silverstein for helpful discussions. We are especially thankful to
John McGreevy for many valuable discussions. Much of this work was
carried out at the Kavli Institute for Theoretical Physics at UCSB,
during the "String Phenomenology" workshop. We would also like to
thank the MIT Center for Theoretical Physics for their generous
hospitality during various parts of this project. A.L. would also
like to thank the theory group at CEA Saclay for their generous
hospitality as this work was completed. This research was supported
in part by the National Science Foundation under Grant No.
PHY99-07949. The research of A.L. and A.S. is further supported in
part by NSF grant PHY-0331516, by DOE Grant No. DE-FG02-92ER40706,
and by an Outstanding Junior Investigator award.

\medskip

\appendix{A}{Definition of $\tilde F$.}

The hypergeometric function ${}_2F_1(\alpha,\beta,\gamma;z)$ has a
branch point at $z = 1$; the branch cut can be taken along the
positive real axis for $z > 1$. In this paper we find that it is
natural to express the correlators in terms of a particular
continuation of $F$ to $z > 1$. We note that, while writing $F$ for
all values of $z$, \DixonQV\ do not discuss such a continuation. The
continuation below is completely consistent with their results.

Recall that an integral definition of the hypergeometric function
(of the first kind) is
\eqn\hypergeometric{\
_2F_1(\alpha,\beta;\gamma;z)={\Gamma(\gamma)\over\Gamma(\beta)
\Gamma(\gamma-\beta)}\int_0^1{t^{\beta-1}(1-t)^{\gamma-\beta-1}\over
(1-tz)^\alpha}dt~,}
where by $a^b$ for $a$ positive and $b$ fractional, we mean the real
positive branch. This integral is unambiguously defined for $|z|<1$.
At $z\sim 1$ it diverges logarithmically while for $z>1$ one must
decide how one continues past the branch point at $z=1$. Our
monodromy integrals are given in term of the sum of continuations
above and below the branch cut. For $\alpha,\beta<1$ (which is
always the case here), that sum can be written as
\eqn\ourhypergeometric{\eqalign{&\ _2\widetilde
F_1(\alpha,\beta;\gamma;z>1)\cr\equiv&{\Gamma(\gamma)\over\Gamma(\beta)
\Gamma(\gamma-\beta)}\[\int_0^{1/z}{t^{\beta-1}(1-t)^{\gamma-\beta-1}\over
({\bf
1-tz})^\alpha}dt+\cos(\pi\alpha)\int_{1/z}^1{t^{\beta-1}(1-t)^{\gamma-\beta-1}\over
({\bf tz-1})^\alpha}dt\]\cr\cr
=&{\Gamma(\gamma)\Gamma(1-\alpha)\over\Gamma(1-\alpha+\beta)
\Gamma(\gamma-\beta)}z^{-\beta}\
_2F_1(1-\gamma+\beta,\beta;1-\alpha+\beta;1/z)\cr
+&\cos(\pi\alpha){\Gamma(\gamma)\Gamma(1-\alpha)\over\Gamma(\gamma+1-\alpha-\beta)
\Gamma(\beta)}{z^{1-\gamma}\over (z-1)^{\alpha+\beta-\gamma}}\
_2F_1(1-\alpha,1-\beta;1-\alpha-\beta+\gamma;1-z)~,}}
and
\eqn\beforebranch{
    \ _2\widetilde F_1(\alpha,\beta;\gamma;z\le 1)=\ _2F_1(\alpha,\beta;\gamma;z\le 1)~.}
Equivalently, for any real value of $z$, one can think of
${}_2\widetilde F_1$ as the limit of
\eqn\abovebelow{\ _2\widetilde F_1(\alpha,\beta;\gamma;z)\equiv
\half\[{}_2 F_1(\alpha,\beta;\gamma;z)+\
_2F_1(\alpha,\beta;\gamma;\bar z)\]~,}
for which $z$ approaches the real line, and the integral in
\hypergeometric\ is evaluated along the real line.

\appendix{B}{Some useful hypergeometric integrals}

In calculating the monodromy integrals in \S2, some useful integrals
for $y\in [0,1]$ are
\eqn\intone{\int_0^ydx[x(1/y-\epsilon_2x)]^{-\nu}(y-x)^{-(1-\nu)}
=y^\nu\Gamma\(1-\nu\)\Gamma\(\nu\)\
_2F_1\(\nu,1-\nu;1;\epsilon_2y^2\)~,}
\eqn\inttwo{\int_{\epsilon_2/y}^ydx[x(1/y-\epsilon_2x)]^{-\nu}(x-y)^{-(1-\nu)}
=-y^\nu\Gamma\(1-\nu\)\Gamma\(\nu\)\
_2F_1\(\nu,1-\nu;1;1-\epsilon_2y^2\)~,}
\eqn\intthree{\int_0^ydx[x(1/y-\epsilon_2x)]^{-\nu}(y-x)^{\nu} =
y^{1+\nu}\Gamma\(1-\nu\)\Gamma\(1+\nu\)\
_2F_1\(\nu,1-\nu;2;\epsilon_2y^2\)~,}
\eqn\intfour{\int_{\epsilon_2/y}^ydx[x(1/y-\epsilon_2x)]^{-\nu}(x-y)^{\nu}
=(y-\epsilon_2/y)y^\nu\Gamma\(1-\nu\)\Gamma\(1+\nu\)\
_2F_1\(\nu,1-\nu;2;1-\epsilon_2y^2\)~,}
where $\ _2F_1(\alpha,\beta;\gamma;z)$ for $z<1$ is given in
\hypergeometric. Another useful relation is
\eqn\hyperrelation{\ _2\widetilde
F_1(\alpha,\beta;\gamma+1;z)={\gamma\over(\gamma-\alpha)(\gamma-\beta)}
(1-z)^{\gamma-\alpha-\beta+1}\d_z\ _2\widetilde
F_1(\gamma-\alpha,\gamma-\beta;\gamma;z)~.}
Note that for $\gamma-\alpha-\beta\in \ZZ$, there is no branch
ambiguity in \hyperrelation.

\bigskip
\listrefs
\end